\documentclass[aps,twocolumn,showkeys,showpacs,preprintnumbers,prd,superscriptaddress,nofootinbib,10pt]{revtex4-1}
\usepackage{graphicx,epsf,bm,amsmath,amsfonts,amssymb,epstopdf,natbib,hyperref,color,appendix,verbatim,multirow,bm,diagbox}
\hypersetup{colorlinks=true,urlcolor=blue,citecolor=blue,linkcolor=blue,menucolor=blue,anchorcolor=blue,filecolor=blue}
\usepackage[shortlabels]{enumitem}

\newcolumntype{?}{!{\vrule width 3pt}}

\date{\today}

\makeatletter
\newcommand{\fmarki}{\ensuremath{\alpha}}
\newcommand{\fmarkii}{\ensuremath{\beta}}
\newcommand{\fmarkiii}{\ensuremath{\gamma}}
\newcommand{\fmarkiv}{\ensuremath{\delta}}
\newcommand{\fmarkv}{\ensuremath{\epsilon}}
\newcommand{\fmarkvi}{\ensuremath{\zeta}}
\newcommand{\fmarkvii}{\ensuremath{\eta}}
\newcommand{\fmarkviii}{\ensuremath{\theta}}
\newcommand{\fmarkix}{\ensuremath{\iota}}
                
\def\@fnsymbol#1{{\ifcase#1\or \fmarki\or \fmarkii\or \fmarkiii\or \fmarkiv\or \fmarkv\or \fmarkvi\or \fmarkvii\or \fmarkviii\or \fmarkix\or \else\@ctrerr\fi}}
\makeatother

\begin{document}

\title{Neutrino cosmology after DESI: tightest mass upper limits, preference for the normal ordering, and tension with terrestrial observations}

\author{Jun-Qian Jiang}
\email{jiangjunqian21@mails.ucas.ac.cn\\junqian.jiang@studenti.unitn.it}
\affiliation{School of Physical Sciences, University of Chinese Academy of Sciences, No.19(A) Yuquan Road, Beijing 100049, China}
\affiliation{Department of Physics, University of Trento, Via Sommarive 14, 38123 Povo (TN), Italy}

\author{William Giar\`{e}}
\email{w.giare@sheffield.ac.uk}
\affiliation{School of Mathematics and Statistics, University of Sheffield, Hounsfield Road, Sheffield S3 7RH, United Kingdom \looseness=-1}

\author{Stefano Gariazzo}
\email{stefano.gariazzo@unito.it}
\affiliation{Instituto de Fisica Teorica (IFT), Universidad Autonoma de Madrid-CSIC, C/ Nicol\'{a}s Cabrera 13-15, E-28049 Madrid, Spain \looseness=-1}
\affiliation{Department of Physics, University of Turin, via P.\ Giuria 1, 10125 Turin (TO), Italy \looseness=-1}
\affiliation{Istituto Nazionale di Fisica Nucleare (INFN), Sezione di Torino, via P.\ Giuria 1, 10125 Turin (TO), Italy}

\author{Maria Giovanna Dainotti}
\email{maria.dainotti@nao.ac.jp}
\affiliation{Division of Science, National Astronomical Observatory of Japan, 2-21-1 Osawa, Mitaka, Tokyo 181-8588, Japan \looseness=-1}
\affiliation{Graduate University for Advanced Studies (SOKENDAI), Shonankokusaimura, Hayama, Kanagawa 240-0115, Japan \looseness=-1}
\affiliation{Space Science Institute, 4765 Walnut Street Suite B, Boulder, CO 80301, USA \looseness=-1}
\affiliation{Nevada Center for Astrophysics, University of Nevada, 4505 Maryland Parkway, Las Vegas, NV 89154, USA \looseness=-1}

\author{Eleonora Di Valentino}
\email{e.divalentino@sheffield.ac.uk}
\affiliation{School of Mathematics and Statistics, University of Sheffield, Hounsfield Road, Sheffield S3 7RH, United Kingdom \looseness=-1}

\author{Olga Mena}
\email{omena@ific.uv.es}
\affiliation{Instituto de F\'{i}sica Corpuscular (IFIC), University of Valencia-CSIS, E-46980, Valencia, Spain \looseness=-1}

\author{Davide Pedrotti}
\email{davide.pedrotti-1@unitn.it}
\affiliation{Department of Physics, University of Trento, Via Sommarive 14, 38123 Povo (TN), Italy}
\affiliation{Trento Institute for Fundamental Physics and Applications (TIFPA)-INFN, Via Sommarive 14, 38123 Povo (TN), Italy}

\author{Simony Santos da Costa}
\email{simony.santosdacosta@unitn.it}
\affiliation{Department of Physics, University of Trento, Via Sommarive 14, 38123 Povo (TN), Italy}
\affiliation{Trento Institute for Fundamental Physics and Applications (TIFPA)-INFN, Via Sommarive 14, 38123 Povo (TN), Italy}

\author{Sunny Vagnozzi}
\email{sunny.vagnozzi@unitn.it}
\affiliation{Department of Physics, University of Trento, Via Sommarive 14, 38123 Povo (TN), Italy}
\affiliation{Trento Institute for Fundamental Physics and Applications (TIFPA)-INFN, Via Sommarive 14, 38123 Povo (TN), Italy}

\begin{abstract}
\noindent The recent DESI Baryon Acoustic Oscillation measurements have led to tight upper limits on the neutrino mass sum, potentially in tension with oscillation constraints requiring $\sum m_{\nu} \gtrsim 0.06\,{\text{eV}}$. Under the physically motivated assumption of positive $\sum m_{\nu}$, we study the extent to which these limits are tightened by adding other available cosmological probes, and robustly quantify the preference for the normal mass ordering over the inverted one, as well as the tension between cosmological and terrestrial data. Combining DESI data with Cosmic Microwave Background measurements and several late-time background probes, the tightest $2\sigma$ limit we find without including a local $H_0$ prior is $\sum m_{\nu}<0.05\,{\text{eV}}$. This leads to a strong preference for the normal ordering, with Bayes factor relative to the inverted one of $46.5$. Depending on the dataset combination and tension metric adopted, we quantify the tension between cosmological and terrestrial observations as ranging between $2.5\sigma$ and $5\sigma$. These results are strenghtened when allowing for a time-varying dark energy component with equation of state lying in the physically motivated non-phantom regime, $w(z) \geq -1$, highlighting an interesting synergy between the nature of dark energy and laboratory probes of the mass ordering. If these tensions persist and cannot be attributed to systematics, either or both standard neutrino (particle) physics or the underlying cosmological model will have to be questioned.
\end{abstract}
\maketitle

\section{Introduction}
\label{sec:introduction}

The discovery of neutrino flavour oscillations proves that neutrinos have mass, in contrast to the original assumption in the Standard Model of Particle Physics (SM). At the time of writing, this represents the only direct evidence for new physics beyond the SM. Determining neutrino properties with high fidelity is therefore of paramount importance in the search for new physics, to which neutrino masses represent one of the most important and accessible portals. Under the assumption that neutrinos are stable over the lifetime of the Universe, they leave various footprints throughout cosmic evolution, at the level of both background expansion and growth of structure. This makes cosmological observations potentially one of the cleanest probes of the sum of the neutrino masses $\sum m_{\nu}$, especially when combining Cosmic Microwave Background (CMB) data with measurements of the late-time expansion history, for instance from Baryon Acoustic Oscillations (BAO)~\cite{Lesgourgues:2012uu,Lattanzi:2017ubx,Vagnozzi:2019utt,Gerbino:2022nvz,DiValentino:2024xsv}. In fact, over the past years various classes of cosmological observations have been used to place upper limits on the masses of active and sterile neutrinos (see for example Refs.~\cite{Zhang:2015rha,Zhang:2015uhk,Cuesta:2015iho,Giusarma:2016phn,Wang:2016tsz,Zhao:2016ecj,Vagnozzi:2017ovm,Guo:2017hea,Wang:2017htc,Chen:2017ayg,Zhao:2017jma,Nunes:2017xon,Giusarma:2018jei,Guo:2018gyo,RoyChoudhury:2018gay,Vagnozzi:2018pwo,Planck:2018vyg,RoyChoudhury:2018bsd,Feng:2019mym,RoyChoudhury:2019hls,Ivanov:2019pdj,Colas:2019ret,Yang:2019uog,Palanque-Delabrouille:2019iyz,Ivanov:2019hqk,Philcox:2020vvt,Hagstotz:2020ukm,Yang:2020tax,Ballardini:2020iws,Zhang:2020mox,Li:2020gtk,Yang:2020ope,Vagnozzi:2020dfn,Giare:2020vzo,RoyChoudhury:2020dmd,Brinckmann:2020bcn,DiValentino:2021hoh,MoradinezhadDizgah:2021upg,Zhang:2021uyp,Sharma:2022ifr,DEramo:2022nvb,Brieden:2022lsd,Kumar:2022vee,Reeves:2022aoi,Tanseri:2022zfe,RoyChoudhury:2022rva,Yadav:2023qfj,Reeves:2023kjx,Pang:2023joc,Bostan:2023ped,Ghirardini:2024yni,Yadav:2024duq}), with potentially important implications for the neutrino mass ordering as well (see e.g.\ Refs.~\cite{Huang:2015wrx,Hannestad:2016fog,Xu:2016ddc,Gerbino:2016ehw,Li:2017iur,Yang:2017amu,Simpson:2017qvj,Schwetz:2017fey,Long:2017dru,Gariazzo:2018pei,Heavens:2018adv,Mahony:2019fyb,Hergt:2021qlh,Jimenez:2022dkn,Gariazzo:2022ahe}). We recall that neutrino oscillation experiments set the lower limit $\sum m_{\nu} \gtrsim 0.1\,{\text{eV}}$ within the inverted ordering (IO), and $\sum m_{\nu} \gtrsim 0.06\,{\text{eV}}$ within the normal ordering (NO)~\cite{deSalas:2020pgw,Esteban:2020cvm}. Therefore, sufficiently tight upper limits on $\sum m_{\nu}$ will have the effect of disfavoring the IO relative to the NO.

Cosmological constraints on $\sum m_{\nu}$ are of course model-dependent, and most upper limits on $\sum m_{\nu}$ are in fact derived within a minimal 7-parameter $\Lambda$CDM+$\sum m_{\nu}$ cosmological model. Nevertheless, one can derive Bayesian model-marginalized constraints which only (or mostly) depend on the adopted dataset rather than on the assumed model, as shown in recent work by some of us~\cite{Gariazzo:2018meg,DiValentino:2022njd}. It is the case that upper limits on $\sum m_{\nu}$ usually weaken when considering extended cosmological models, especially within those allowing for more freedom at low redshifts, for instance when changing the spatial curvature of the Universe or altering the properties of dark energy (DE) from those of a cosmological constant with equation of state (EoS) $w=-1$. There is, however, a physically interesting counterexample: when imposing that the (evolving) DE component lies in the \textit{quintessence}-like regime $w(z) \geq -1$, i.e.\ should be non-\textit{phantom} (as expected within the simplest, most physically motivated scalar field models for DE, with the phantom regime where the null energy condition is violated corresponding to $w<-1$), the upper limits on $\sum m_{\nu}$ actually become even tighter than those obtained within $\Lambda$CDM, as first explained by some of us in Ref.~\cite{Vagnozzi:2018jhn} and later confirmed by several independent works (e.g.\ Refs.~\cite{RoyChoudhury:2018vnm,Berghaus:2024kra,Green:2024xbb}). This highlights a potentially interesting interplay between the nature of DE, and laboratory experiments aimed (among other things) at determining the neutrino mass ordering (e.g.\ long-baseline experiments such as DUNE~\cite{DUNE:2020jqi} and Hyper-Kamiokande~\cite{Hyper-Kamiokande:2015xww}).

Pre-2024 cosmological constraints on $\sum m_{\nu}$ within the minimal $\Lambda$CDM+$\sum m_{\nu}$ model are particularly tight, ranging between 95\% confidence level (C.L.) upper limits of $\sum m_{\nu}<0.12\,{\text{eV}}$~\cite{Vagnozzi:2017ovm,Planck:2018vyg} down to the tightest $\sum m_{\nu}<0.09\,{\text{eV}}$~\cite{Palanque-Delabrouille:2019iyz,DiValentino:2021hoh,DiValentino:2022njd} and $\sum m_{\nu}<0.08\,{\text{eV}}$~\cite{Brieden:2022lsd}. These constraints are obviously placing the IO under some tension, with the exact values for the NO vs IO odds varying depending on a number of underlying assumptions. As these upper limits keep getting tighter in the absence of a detection of non-zero $\sum m_{\nu}$, another even more puzzling tension may emerge between cosmological upper limits on $\sum m_{\nu}$ and the lower limit of $0.06\,{\text{eV}}$ set by terrestrial experiments. These discussions, until some time ago somewhat academical in nature, have now become all the more urgent in light of BAO measurements from the Dark Energy Spectroscopic Instrument (DESI)~\cite{DESI:2016fyo,DESI:2016igz} first year of observations~\cite{DESI:2024mwx}, whose implications for fundamental physics have been explored in several works (see e.g.\ Refs.~\cite{Tada:2024znt,Yin:2024hba,Wang:2024hks,Luongo:2024fww,Cortes:2024lgw,Clifton:2024mdy,Colgain:2024xqj,Carloni:2024zpl,Wang:2024rjd,Allali:2024cji,Giare:2024smz,Wang:2024dka,Gomez-Valent:2024tdb,Yang:2024kdo,Park:2024jns,Escamilla-Rivera:2024sae,Wang:2024pui,Shlivko:2024llw,DESI:2024aqx,Dinda:2024kjf,Bousis:2024rnb,Seto:2024cgo,Favale:2024sdq,Croker:2024jfg,DESI:2024kob,Bhattacharya:2024hep,Ramadan:2024kmn,Mukherjee:2024ryz,Pogosian:2024ykm,Roy:2024kni,Jia:2024wix,Wang:2024hwd,Heckman:2024apk,Gialamas:2024lyw,Notari:2024rti,Lynch:2024hzh,Liu:2024txl,Chudaykin:2024gol,Dwivedi:2024okk,Liu:2024gfy,Orchard:2024bve,Patel:2024odo,Hernandez-Almada:2024ost,Pourojaghi:2024tmw,Mukhopadhayay:2024zam,Wang:2024sgo,Li:2024qso,Ye:2024ywg,Giare:2024gpk,Dinda:2024ktd,Jiang:2024xnu,Sohail:2024oki,Escamilla:2024xmz,Sharma:2024mtq,Ghosh:2024kyd,Pourojaghi:2024bxa,Sabogal:2024yha,Escamilla:2024ahl,Pang:2024qyh,Wolf:2024eph,Shao:2024mag,Ghedini:2024mdu,Taule:2024bot,RoyChoudhury:2024wri,Lu:2024hvv,Arjona:2024dsr,Giare:2024ocw,Wolf:2024stt,Wang:2024tjd,Dhawan:2024gqy,Loverde:2024nfi,Alestas:2024eic,Linder:2024rdj,Chan-GyungPark:2024brx,Giani:2024nnv,Menci:2024hop,Alfano:2024fzv,Luongo:2024zhc,Wu:2024faw,Gao:2024ily}). Taken at face value, besides puzzling hints for dynamical DE, the DESI BAO measurements combined with CMB data from the \textit{Planck} satellite and the Atacama Cosmology Telescope (ACT) also set a particularly tight limit on $\sum m_{\nu}<0.072\,{\text{eV}}$ at 95\%~C.L. (with negative neutrino masses in principle preferred by the data)~\cite{DESI:2024mwx}, which is uncomfortably close to scratching the surface of the minimum value allowed by terrestrial oscillations. Such a tension becomes even more evident in the analysis conducted by some of us in Ref.~\cite{Wang:2024hen}, where the addition of extra background probes of the expansion history (e.g.\ cosmic chronometers, galaxy cluster angular diameter distances, and gamma-ray bursts) results in bounds as tight as $\sum m_{\nu}<0.043\,{\text{eV}}$. It goes without saying that the implications of these results for new (cosmological and/or particle) physics are potentially momentous.

The above results raise a number of pressing questions, including but not limited to (see also the recent Refs.~\cite{Du:2024pai,Reboucas:2024smm} for related studies):
\begin{enumerate}[(a)]
\item how far can current data (including cosmological probes beyond the most widely used ones, and other than the ones used in the DESI analysis~\cite{DESI:2024mwx}) go insofar as upper limits on $\sum m_{\nu}$ are concerned?
\item to what extent is the inverted ordering disfavored relative to the normal ordering?
\item what is the level of tension (if any) between cosmological and terrestrial measurements?
\end{enumerate}
It is our goal in this work to address the above questions, not only within the minimal $\Lambda$CDM+$\sum m_{\nu}$ model, but also considering physically motivated time-varying DE models whose EoS is restricted to the non-phantom regime, $w(z) \geq -1$. For what concerns the tension between terrestrial and cosmological experiments -- question (\textit{c}) --, we make use of the methods developed by some of us in Ref.~\cite{Gariazzo:2023joe} to robustly quantify the level of disagreement. As a quick appetizer of our results, we find that within the most aggressive dataset combination (which includes CMB data from \textit{Planck} and ACT, DESI BAO data, galaxy cluster angular diameter distance measurements, and a local prior on $H_0$) $\sum m_{\nu}<0.042\,{\text{eV}}$ ($0.041\,{\text{eV}}$) for the $\Lambda$CDM+$\sum m_{\nu}$ (+$w(z) \geq -1$) model, leading to a Bayes factor of $72.6$ ($109.2$) for the NO relative to the IO, whereas the level of tension with terrestrial experiments ranges between $4$ and $5\sigma$ depending on the adopted tension metric. Taken at face value, our results appear to signal the end of the line for the IO, but also a somewhat concerning tension between cosmology and terrestrial experiments.

The rest of this paper is then organized as follows. We describe the adopted datasets and methodology in Sec.~\ref{sec:data}. Our results, in particular our upper limits on $\sum m_{\nu}$, Bayes factors for normal versus inverted ordering, and quantification of the tension between cosmology and terrestrial experiments, are discussed in Sec.~\ref{sec:results}. Finally, we draw concluding remarks in Sec.~\ref{sec:conclusions}. A brief discussion of how our results change when the dark energy component is allowed to explore the phantom regime is carried out in Appendix~\ref{sec:appendix}. The level of internal consistency between our cosmological probes is discussed in Appendix~\ref{sec:appendixb}. The impact of using the \textit{Planck} PR4 likelihoods in place of their PR3 counterparts is instead assessed in Appendix~\ref{sec:appendixc}. In Appendix~\ref{sec:appendixd} we study the impact of treating the SH0ES information as a prior on $H_0$ rather than on $M_B$.

\section{Datasets and methodology}
\label{sec:data}

\subsection{Models}
\label{subsec:models}

The baseline model we consider is the 7-parameter $\Lambda$CDM+$\sum m_{\nu}$ model where, in addition to the 6 standard parameters of $\Lambda$CDM (the acoustic angular scale $\theta_s$, the physical baryon and cold dark matter densities $\omega_b$ and $\omega_c$, the amplitude and tilt of the primordial scalar power spectrum $A_s$ and $n_s$, and the optical depth to reionization $\tau$), we vary the sum of the neutrino masses $\sum m_{\nu}$. In addition, we consider a model with more freedom in the DE sector, where the DE EoS is parametrized as per the widely used 2-parameter Chevallier-Polarski-Linder (CPL) parametrization~\cite{Chevallier:2000qy,Linder:2002et}:
\begin{eqnarray}
w(z) = w_0+w_a\frac{z}{1+z}\,,
\label{eq:cpl}
\end{eqnarray}
where $z$ indicates redshift, $w_0$ is the present-day DE EoS, and $w_a$ parametrizes the redshift evolution of $w$. Eq.~(\ref{eq:cpl}) is clearly a truncated Taylor expansion in the scale factor around $a\simeq a_0\equiv 1$, and benefits of such a parametrization, including its direct connection to physically interesting DE models, have been widely discussed in the literature~\cite{Linder:2007wa,Linder:2008pp}.~\footnote{For a recent summary on this point, see e.g.\ the text below Eq.~(2.1) in Ref.~\cite{Adil:2023ara}. See instead Refs.~\cite{Efstathiou:1999tm,Jassal:2004ej,Gong:2005de,Barboza:2008rh,Ma:2011nc,Pantazis:2016nky,Pan:2017zoh,Yang:2018qmz,Visinelli:2019qqu,Adil:2023exv} for examples of other dynamical DE parametrization in the literature.}

If $w_0$ and $w_a$ are allowed to take arbitrary values (of course with $w_0<-1/3$ in order to ensure cosmic acceleration today), the DE EoS may cross the phantom divide $w=-1$ at some point during the cosmic evolution. Such a behaviour, however, is not possible within the simplest and arguably best motivated \textit{quintessence} DE models, based on a single scalar field with canonical kinetic term, minimally coupled to gravity, and in the absence of higher-derivative operators~\cite{Wetterich:1987fm,Ratra:1987rm,Wetterich:1994bg,Caldwell:1997ii,Sahni:2002kh}.~\footnote{Nevertheless, at face value these models appear to be disfavored observationally, as they worsen the Hubble tension~\cite{OColgain:2018czj,Colgain:2019joh,Vagnozzi:2019ezj,Banerjee:2020xcn,Heisenberg:2022gqk,Lee:2022cyh,Escamilla:2023oce}.} Indeed, a phantom behaviour is considered to be problematic from the theoretical point of view, given the implied violation of the null energy condition, whereas from the cosmological point of view it typically results in a so-called ``\textit{Big Rip}''~\cite{Caldwell:2003vq} (see also Refs.~\cite{Nojiri:2005sx,Astashenok:2012tv,Odintsov:2015zza,Odintsov:2018zai,Trivedi:2023zlf}). For this reason, we pay special consideration to the theoretically well-motivated case where $w(z)$ in Eq.~(\ref{eq:cpl}) is constrained to $w(z) \geq -1$ at all redshifts. Noting that $w(z)$ in the CPL parametrization is a monotonic function of $z$ (whether it is increasing or decreasing depends on the sign of $w_a$) which goes from $w_0$ at $z=0$ to $w_0+w_a$ as $z \to \infty$, we see that it is sufficient to impose that $w_0 \geq -1$ \textit{and} $w_0+w_a \geq -1$ to ensure $w(z) \geq -1$ at all times~\cite{Vagnozzi:2018jhn}. Following Ref.~\cite{Vagnozzi:2018jhn}, where this particular region of CPL parameter space was first studied in detail by some of us, we denote by ``NPDDE'' (standing for ``non-phantom dynamical dark energy'') the 8-parameter model where, in addition to the 6 $\Lambda$CDM parameters, we vary $w_0$ and $w_a$ subject to the physically motivated constraints $w_0 \geq -1 \wedge w_0+w_a \geq -1$.~\footnote{It is worth noting that the choice of such a (physically motivated) prior becomes somewhat crucial when analyzing DESI BAO measurements. When $w_0$ and $w_a$ are allowed to vary freely and a phantom crossing is possible, the combined analysis of DESI and \textit{Planck} CMB data shows a preference for a dynamical DE component with a present-day quintessence-like EoS that crossed the phantom divide in the past, thus violating our prior assumption $w_0 + w_a \geq -1$. Such a preference can exceed the $3\sigma$ level when SNeIa measurements are included~\cite{DESI:2024uvr}. For further discussions on this point see Appendix~\ref{sec:appendix}.}

Accordingly, we also consider the 9-parameter NPDDE+$\sum m_{\nu}$ model, where $\sum m_{\nu}$ is varied alongside the aforementioned 8 parameters. As alluded to in the Introduction and discussed in detail in Ref.~\cite{Vagnozzi:2018jhn} (to which we refer the reader for a complete explanation), upper limits on $\sum m_{\nu}$ within the NPDDE+$\sum m_{\nu}$ model are tighter than those obtained within the $\Lambda$CDM+$\sum m_{\nu}$ one, in spite of the 2 additional parameters which are marginalized on.~\footnote{At first glance, such a result may appear in contradiction with the Cram\'{e}r-Rao bound. As explicitly argued in Ref.~\cite{Green:2024xbb}, this is actually not the case, since the inclusion of physical priors which restrict the parameter space, such as $\sum m_{\nu} \geq 0$, can make a given parameter appear to be more tightly constrained due to shifts in its central value (in this case $\sum m_{\nu}$ shifting towards more negative values) rather than a decrease in its variance (which actually increases, although this is not obvious unless one removes the physical prior).} This represents perhaps the most important and physically relevant counterexample to the standard lore according to which upper limits on $\sum m_{\nu}$ typically degrade in extended models. For completeness, we also consider the $w_0w_a$CDM+$\sum m_{\nu}$ model, where $w(z)$ is still modeled following Eq.~(\ref{eq:cpl}), but $w_0$ and $w_a$ are not subject to the NPDDE constraints and the DE EoS is therefore free to cross the phantom divide -- however, since the ensuing constraints are not of direct interest to this work, we briefly discuss them in Appendix~\ref{sec:appendix}.

We set wide, flat priors on all cosmological parameters, verifying a posteriori that our posteriors are not affected by the choice of lower and upper prior boundaries (except for the case of the physically motivated NPDDE priors on $w_0$ and $w_0+w_a$, see also footnote~3). However, a few comments are in order for what concerns our treatment of $\sum m_{\nu}$. Firstly, we model the neutrino mass spectrum as consisting of three degenerate mass eigenstates, each with mass $\sum m_{\nu}/3$. While this does not account for the mass splittings inferred from oscillation experiments, such an approximation has extensively been shown to be sufficiently accurate for the purposes of current data, which are only sensitive to the total neutrino mass $\sum m_{\nu}$, but not to how this is distributed among the three eigenstates~\cite{Giusarma:2016phn,Gerbino:2016sgw,Archidiacono:2016lnv,CORE:2016npo,Archidiacono:2020dvx}. In other words, adopting the degenerate approximation does not lead to biases in the inferred cosmological parameters compared to the case where the actual mass splittings are modeled. We recall in fact that the mass splittings are not relevant for background quantities, which are the main quantities of interest to this work. Moreover, their impact on the growth of structure and the CMB lensing signal is negligible, as shown in several earlier works~\cite{Hall:2012kg,CORE:2016npo,Archidiacono:2020dvx}.

\begin{table*}[!t]
\scalebox{1.0}{
\begin{tabular}{|c?c|c|c|}
\hline
\textbf{Model} & \texttt{\#} parameters & Free parameters & Priors \\
\hline \hline
$\Lambda$CDM+$\sum m_{\nu}$ & 7 & $\omega_b$, $\omega_c$, $\theta_s$, $A_s$, $n_s$, $\tau$, $\sum m_{\nu}$ & $\sum m_{\nu} \geq 0\,{\text{eV}}$ \\ \hline
NPDDE+$\sum m_{\nu}$ & 9 & $\omega_b$, $\omega_c$, $\theta_s$, $A_s$, $n_s$, $\tau$, $w_0$, $w_a$, $\sum m_{\nu}$ & $\sum m_{\nu} \geq 0\,{\text{eV}}\,, \quad w_0 \geq -1\,, \quad w_0+w_a \geq -1$ \\ \hline
$w_0w_a$CDM+$\sum m_{\nu}$ & 9 & $\omega_b$, $\omega_c$, $\theta_s$, $A_s$, $n_s$, $\tau$, $w_0$, $w_a$, $\sum m_{\nu}$ & $\sum m_{\nu} \geq 0\,{\text{eV}}$ \\ \hline
\end{tabular}}
\caption{Summary of the 3 cosmological models considered in this work.}
\label{tab:models}
\end{table*}

In addition, following standard practice in the field, the prior we impose on the sum of the neutrino masses is $\sum m_{\nu} \geq 0\,{\text{eV}}$, thereby not accounting for information from oscillation experiments which require $\sum m_{\nu} \gtrsim 0.06\,{\text{eV}}$. There are various good reasons for adopting this choice, the most important one being that the positivity of $\sum m_{\nu}$ is the only genuinely a priori physical information in the problem, therefore ensuring that the resulting bound on $\sum m_{\nu}$ relies exclusively on cosmological data. Moreover, such a choice allows for an interesting consistency test of cosmological models (or unaccounted for systematics), should the resulting upper limit on $\sum m_{\nu}$ be in tension with the lower limit set by oscillation experiments~\cite{Gariazzo:2023joe}. We refer the reader to Ref.~\cite{Vagnozzi:2018jhn} for further discussions on the rationale behind this choice and its merits. Finally, although some works analyzing the impact of DESI data on $\sum m_{\nu}$ constraints have explored the impact of allowing for negative values of $\sum m_{\nu}$~\cite{Craig:2024tky,Green:2024xbb,Elbers:2024sha,Noriega:2024lzo,Naredo-Tuero:2024sgf}, here we shall not adopt such a phenomenological choice, once more because the only genuine a priori physical information is the positivity of $\sum m_{\nu}$. Key details (free parameters and priors thereon) of the 3 models we study are summarized in Tab.~\ref{tab:models}. For all cases we fix the effective number of relativistic species to $N_{\text{eff}}=3.044$, in agreement with some of the latest determinations~\cite{Akita:2020szl,Froustey:2020mcq,Bennett:2020zkv,Cielo:2023bqp,Drewes:2024wbw}.

\subsection{Datasets}
\label{subsec:data}

The models in question are then confronted against a number of state-of-the-art cosmological observations. We first consider a set of cosmological observations which can be considered somewhat standard.
\begin{itemize}
\item \textit{Cosmic Microwave Background} -- we adopt the same CMB dataset used in the official DESI analysis~\cite{DESI:2024mwx}, which combines the \textit{Planck} PR3 temperature and polarization anisotropy (TTTEEE) likelihoods~\cite{Planck:2018vyg}, the PR4 likelihood for the lensing power spectrum reconstructed from the temperature 4-point function, and the ACT DR6 lensing power spectrum likelihood~\cite{ACT:2023kun}. We also considered the ``extended'' version of the ACT lensing likelihood (deemed to be trustworthy as explained in Ref.~\cite{ACT:2023kun}), which extends the baseline range of multipoles from $40<\ell<763$ to $40<\ell<1300$.~\footnote{A few months ago, the ACT likelihood versions were updated by the collaboration from \texttt{v1.1} to \texttt{v1.2}. The DESI collaboration adopted \texttt{v1.1} in their analyses, and for consistency our baseline analyses also adopt this version. Nevertheless, in what follows we will also very briefly discuss the impact of switching to \texttt{v1.2}.} The impact of using the PR4 \texttt{LoLLiPoP} and \texttt{HiLLiPoP} likelihoods in place of their PR3 counterparts has been explored in Ref.~\cite{Allali:2024aiv}, finding that this results in slightly weakened limits. We also check the impact of using these likelihoods for a subset of our results in Appendix~\ref{sec:appendixc}, see Tab.~\ref{tab:PR4}, confirming these findings.
\item \textit{Baryon Acoustic Oscillations} -- we use the DESI BAO measurements~\cite{DESI:2024mwx} in the range $0.1<z<4.16$, based on observations of the clustering of the Bright Galaxy Sample (BGS), the Luminous Red Galaxy Sample (LRG), the Emission Line Galaxy (ELG) Sample and the combined LRG+ELG sample, quasars, and the Lyman-$\alpha$ forest. This represents our baseline BAO dataset. In an extended setting, we also consider the combination with the earlier SDSS BAO measurements, following the conservative approach discussed in Section~3.3 and Appendix~A of Ref.~\cite{DESI:2024mwx}, to which we refer the reader for more detailed discussions.~\footnote{Specifically, the SDSS measurements at $z=0.15$ (MGS), $0.38$, and $0.51$ (two lowest redshift bins for BOSS galaxies) replace the DESI BGS and LRG samples, whereas the combined DESI+SDSS sample is used for Lyman-$\alpha$ forest BAO.}
\item \textit{Type Ia Supernovae} -- we make use of distance moduli measurements from the (uncalibrated) \textit{PantheonPlus} Type Ia Supernovae (SNeIa) sample in the redshift range $0.01<z<2.26$~\cite{Scolnic:2021amr}.
\item \textit{SH0ES} -- in our most ``aggressive'' (and less conservative) dataset combinations we consider a Gaussian prior on the Hubble constant $H_0=(73.04 \pm 1.04)\,{\text{km}}/{\text{s}}/{\text{Mpc}}$, itself motivated by the Cepheid-calibrated SNeIa distance ladder measurement presented in Ref.~\cite{Riess:2021jrx}. Since our DE models are smooth and do not feature abrupt transitions at extremely low redshifts, we expect that including the SH0ES information as a prior on $H_0$ rather than on $M_B$, or using the full joint \textit{PantheonPlus}+SH0ES likelihood, should have little impact on our results. We confirm this in Appendix~\ref{sec:appendixd}, see Tab.~\ref{tab:H0_prior}.
\end{itemize}
In addition, we also make use of a number of less widely adopted cosmological datasets.
\begin{itemize}
\item \textit{Cosmic Chronometers} -- measurements of the expansion rate $H(z)$ from so-called cosmic chronometers (CC), i.e.\ the differential ages of massive, early-time, passively-evolving galaxies~\cite{Jimenez:2001gg}. We use 15 data points compiled in Refs.~\cite{Moresco:2012by,Moresco:2015cya,Moresco:2016mzx}. These are measurements for which a full estimate of non-diagonal covariance terms, including contributions from systematics, is available. These contributions have been estimated following the methodology proposed in Refs.~\cite{Moresco:2018xdr,Moresco:2020fbm}, making these measurements safe against the concerns raised in Ref.~\cite{Kjerrgren:2021zuo}. The CC measurements constrain $H(z)$ in the range $0.1791<z<1.965$.
\item \textit{Galaxy Clusters} -- combining X-Ray and Sunyaev-Zeldovich (XSZ) observations of galaxy clusters from Refs.~\cite{Mason:2001ghw,Reese:2002sh}, and using an isothermal elliptical $\beta$ model to model the elliptical surface brightness of the latter, Ref.~\cite{DeFilippis:2005hx} derived angular diameter distance measurements for 25 clusters in the range $0.023<z<0.784$, which we make use of here.
\item \textit{Gamma-Ray Bursts} -- we adopt the \textit{Platinum Sample} of Gamma-Ray Bursts (GRB), used as distance indicators to constrain the luminosity distance-redshift evolution in the range $0.553<z<5$~\cite{Dainotti:2020azn}. This sample has well-defined lightcurves, and improves upon the earlier \textit{Gold Sample}~\cite{Dainotti:2016iqn,Dainotti:2017fem,Dainotti:2020azn,Dainotti:2020jkj,Dainotti:2022jzv}, and for which the plateau emission is fit with the Ref.~\cite{Willingale:2006zh} function if the following quality criteria are satisfied: the angle characterizing the plateau emission should be $<41^{\circ}$~\citep{Dainotti:2016iqn}, there should be no flares and gaps in the plateau emission which at its beginning should have at least 5 data points, and the plateau duration should be $> $500$\,{\text{s}}$. The distribution variables for the sample have been checked against the full distribution in Ref.~\cite{Dainotti:2022mto}, whereas the likelihood for the observed distance moduli has been constructed following Ref.~\cite{Dainotti:2022wli,Dainotti:2022rea,Dainotti:2022ked,Dainotti:2023bwq,Adil:2024miw}. Without loss of generality, we treat the redshift evolution correction by fixing the evolutionary coefficients to their mean values, similarly to Ref.~\cite{Dainotti:2022ked}.
\end{itemize}
While less widely adopted, the CC, XSZ, and GRB datasets can be particularly important for our work due to the fact that they constrain the late-time background expansion of the Universe. This is crucial to further break the geometrical degeneracy once combined with CMB measurements, improving the determination of $\Omega_m$ and $H_0$ alongside the DE parameters. For this reason, these measurements play an important role when attempting to constrain $\sum m_{\nu}$, given the important correlations between $\sum m_{\nu}$ and the aforementioned parameters. For detailed discussions on the impact of $\sum m_{\nu}$ on cosmological observables, as well as degeneracies with other parameters, we refer the reader to detailed reviews on the subject~\cite{Lesgourgues:2006nd,Lattanzi:2017ubx,Gerbino:2022nvz}. We note that the CC, XSZ, and GRB datasets have been used earlier by some of us in Ref.~\cite{Wang:2024hen} to improve upper limits on neutrino masses. With respect to this earlier work, we consider several other dataset combinations, study the stability of our results against likelihood settings, and quantify the tension with terrestrial experiments. Moreover, our GRB dataset is corrected for selection biases and redshift evolution, a non-trivial often overlooked aspect. It has been shown in Ref.~\cite{Dainotti:2013cta} that a deviation of 5$\sigma$ both towards higher or lower values in the intrinsic slope of the luminosity-break time relationship, key to the use of GRBs as standard candles, can lead to parameter biases as large as $13\%$. This marks the necessity to correct for these effects, naturally accounted for in the GRB dataset we use.

\begin{figure}[!ht]
\centering
\includegraphics[width=0.75\linewidth]{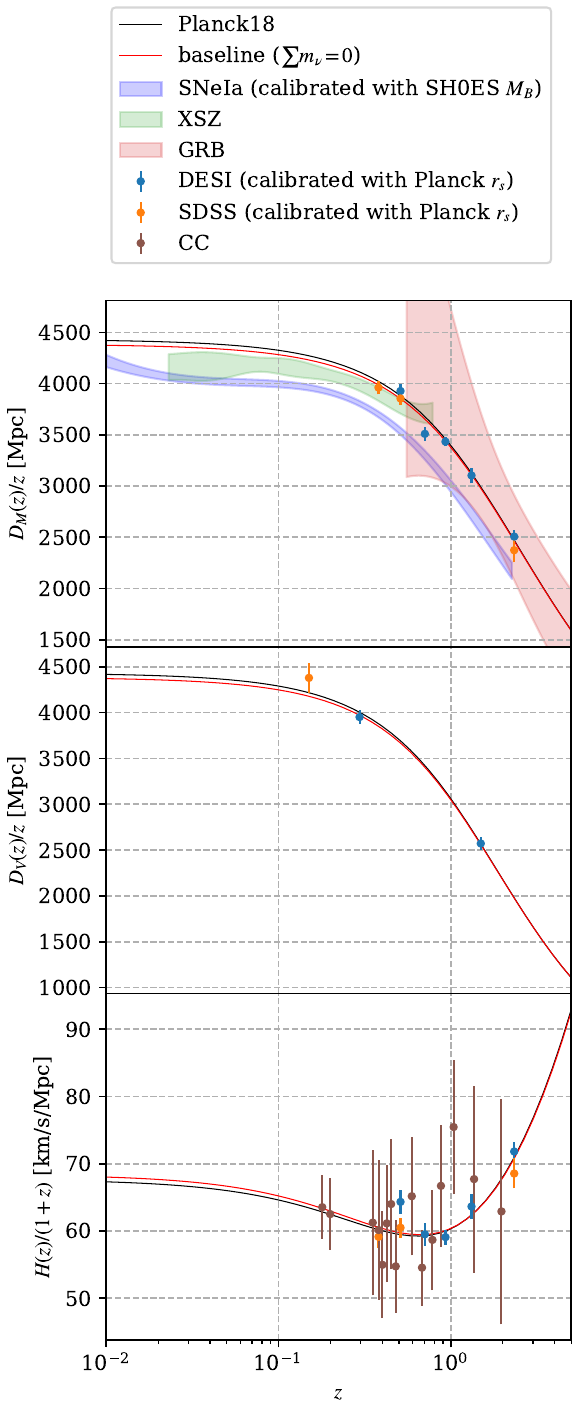}
\caption{Datasets used in this work (see color coding and labels). Each panel corresponds to a specific background quantity: $D_M(z)/z$ (upper panel), $D_V(z)/z$ (intermediate panel), and $H(z)/(1+z)$ (lower panel). On each panel we plot the evolution of the same functions within the $\Lambda$CDM model with parameters given by a fit to \textit{Planck} data alone (black curves), assuming one massive neutrino with $\sum m_{\nu}=0.06\,{\text{eV}}$, and the same evolution given by a fit to our baseline dataset combination (red curves), assuming $\sum m_{\nu}=0\,{\text{eV}}$. SNeIa and BAO are calibrated using the SH0ES prior on the SNeIa absolute magnitude $M_B=-19.253 \pm 0.027$ and \textit{Planck} prior on the sound horizon $r_d = (147.09 \pm 0.26)\,{\text{Mpc}}$. For SNeIa, XSZ, and GRB, given the large amount of points and/or large uncertainties of some of these, we do not display individual points, but reconstruct the corresponding $D_M(z)/z$ using Gaussian Process regression (colored bands indicate $68\%$ credible intervals for the reconstructed functions).}
\label{fig:distance}
\end{figure}

The adopted datasets are plotted in Fig.~\ref{fig:distance}, where each panel corresponds to a specific background quantity: $D_M(z)/z$, with $D_M(z)$ the transverse comoving distance (upper panel), $D_V(z)/z$ (intermediate panel), with $D_V(z)$ the volume-averaged distance, and $H(z)/(1+z)$ (lower panel). On each panel we plot the evolution of the same functions within the $\Lambda$CDM model with parameters determined by a fit to \textit{Planck} data alone (black curves), assuming one massive neutrino with $\sum m_{\nu}=0.06\,{\text{eV}}$, and the same $\Lambda$CDM evolution determined from a fit to our baseline dataset combination (red curves), assuming $\sum m_{\nu}=0\,{\text{eV}}$. We note that the SNeIa and BAO datasets have been calibrated using the SH0ES prior on the SNeIa absolute magnitude $M_B=-19.253 \pm 0.027$ and \textit{Planck} prior on the sound horizon $r_d = (147.09 \pm 0.26)\,{\text{Mpc}}$ respectively. Finally, for what concerns $D_M(z)/z$ as measured by SNeIa, XSZ, and GRB, given the large amount of datapoints and/or relatively large uncertainties of some of the points, we do not display the individual datapoints, but reconstruct the corresponding $D_M(z)/z$ using Gaussian Process regression, following the methodology of Ref.~\cite{Jiang:2024xnu}. Inspecting the $D_M(z)/z$ panel, one is brought to the following expectations: \textit{a)} the SH0ES prior is in significant tension with the other probes and pushes towards the $\sum m_{\nu}<0\,{\text{eV}}$ region (which, given our positive $\sum m_{\nu}$ prior, implies tighter limits on the latter), and should therefore lead to artificially tighter constraints on $\sum m_{\nu}$; \textit{b)} the XSZ dataset is in $\approx 2\sigma$ (dis)agreement with the other probes, and goes in the direction of pushing towards smaller values of $\sum m_{\nu}$, but not as strongly as the SH0ES prior (and without leading to artificial tension with the other datasets); and \textit{c)} due to its large uncertainties, the GRB dataset is not expect to lead to significant improvements in the constraints on $\sum m_{\nu}$. All three expectations in fact turn out to be explicitly verified in our later analysis (see Sec.~\ref{sec:results} and Appendix~\ref{sec:appendixb}).

\subsection{Methods and tension metrics}
\label{subsec:methods}

To sample the posterior distributions of the cosmological parameters in question we make use of Monte Carlo Markov Chain (MCMC) methods via the cosmological MCMC sampler \texttt{Cobaya}~\cite{Torrado:2020dgo}. Theoretical predictions for the relevant cosmological observables are obtained through the Boltzmann solver \texttt{CAMB}~\cite{Lewis:1999bs}. As recommended by the ACT collaboration, to deal with non-linear corrections we make use of \texttt{HMCode}~\cite{Mead:2016zqy}, while requiring higher numerical accuracy settings at the Boltzmann solver level.~\footnote{The version of \texttt{HMCode} used by the ACT collaboration is the 2016 one~\cite{Mead:2016zqy}. Nevertheless, we note that the newer \texttt{HMCode-2020} is available~\cite{Mead:2020vgs}. In what follows, we briefly discuss the impact of switching between the two versions.} We assess the convergence of our MCMC chains using the Gelman-Rubin $R-1$ parameter~\cite{Gelman:1992zz}, and consider our chains converged when $R-1<0.01$. To reduce sampling errors in the tails of the posterior distributions, we resort to tempered MCMC for cases where the upper limits on $\sum m_{\nu}$ are particularly tight, and the sample counts at $\sum m_{\nu}>0.10\,{\text{eV}}$ are less than $1000$: in this case, we set the \texttt{temperature} flag of \texttt{Cobaya} to a value $t>1$, thereby sampling a power-reduced version of the posterior $p$, which is softened to $p^{1/t}$, allowing for a more precise sampling of the tails of the distributions in question.

In addition to deriving upper limits on $\sum m_{\nu}$, other important goals of our work are to properly quantify the preference for the NO over the IO, as well as the tension between cosmological and terrestrial experiments. For what concerns the former, we quantify this preference by computing the Bayes factor for NO versus IO:
\begin{eqnarray}
B_{\text{NO,IO}} \equiv \frac{\mathcal{Z}_{\rm NO}}{\mathcal{Z}_{\rm IO}} \,,
\label{eq:B}
\end{eqnarray}
where $\mathcal{Z}$ denotes the Bayesian evidence for the model in question. As it is defined, the ratio $B_{\text{NO,IO}}$ then quantifies the Bayesian odds in favor of the NO, with values of $B_{\text{NO,IO}}>1$ denoting a preference for the NO. Finally, if equal prior odds are assumed for the NO and IO, the Bayes factor can be converted into posterior probabilities for the NO and IO, $P_{\text{NO}}=B_{\text{NO,IO}}/(1+B_{\text{NO,IO}})$ and $P_{\text{IO}}=1/(1+B_{\text{NO,IO}})$, obviously with $P_{\text{NO}}+P_{\text{IO}}=1$. The strength of the preference for NO versus IO is qualified using a modified version of the Kass-Raftery scale (essentially with $\log_{10}$ replaced by the natural logarithm)~\cite{Kass:1995loi}, itself a modified version of the Jeffreys scale. On this scale, values of $\ln B_{\rm NO,IO}<0$ indicate a preference for the IO, values of $0 \leq \ln B_{\rm NO,IO}<1$ indicate a weak preference for the NO, values of $1 \leq \ln B_{\rm NO,IO}<3$ indicate a positive preference for the NO, values of $3 \leq \ln B_{\rm NO,IO}<5$ indicate a strong preference for the NO, and values of $\ln B_{\rm NO,IO} \geq 5$ indicate a very strong preference for the NO. See e.g.\ Tab.~2 of Ref.~\cite{Yang:2018euj} or Tab.~1 of Ref.~\cite{Vagnozzi:2019ezj} for examples of the use of this scale in cosmology.

To evaluate the tension between cosmological and terrestrial experiments, we make use of the method developed by some of us in Ref.~\cite{Gariazzo:2023joe}. To this end, on the terrestrial data side, we employ a combination of data from neutrino oscillation experiments, and the KATRIN $\beta$-decay experiment. For what concerns the results of oscillation experiments, we treat them in terms of a Gaussian likelihood on the solar and atmospheric mass-squared splittings, with mean and standard deviations given by the following~\cite{deSalas:2020pgw,Esteban:2020cvm,Capozzi:2021fjo}:~\footnote{Note that the $\Delta\chi^2$ between the NO and IO based on oscillation data does not affect the tension metrics we make use of and is therefore not relevant to our analyses.}
\begin{equation}
\begin{split}
\Delta m^2_{21} &=(7.50\pm 0.21)\times 10^{-5} \,{\rm eV}^2 \,, \\
|\Delta m^2_{31}|&= \left\{
\begin{array}{ll}
(2.550\pm0.025)\times 10^{-3}\, {\rm eV}^2 &\quad \rm (NO) \\
(2.450\pm0.025)\times 10^{-3}\, {\rm eV}^2 &\quad \rm (IO)
\end{array}
\right. \,,
\end{split}
\label{eq:oscillations}
\end{equation}
where $\Delta m^2_{ij} \equiv m_i^2-m_j^2$. The above essentially translate into the previously mentioned lower bounds on $\sum m_{\nu}$ from oscillation experiments for the two different neutrino mass orderings. For what concerns the KATRIN results, we instead adopt a Gaussian likelihood on the effective neutrino mass $m_{\beta}$ as determined by the combination of the first and second mass campaigns, as follows~\cite{KATRIN:2021uub}:~\footnote{Notice that our constraint does not reflect the most recent results by the KATRIN collaboration \cite{Katrin:2024tvg}, which were released at a late stage of the preparation of this work. Our results, however, are almost unaffected by a stronger upper bound from terrestrial experiments, since the tension only involves the lower $\sum m_{\nu}$ limit provided by neutrino oscillations.}
\begin{eqnarray}
m_{\beta}^2 = (0.06 \pm 0.32)\,\text{eV}^2 \,.
\label{eq:mbeta}
\end{eqnarray}
The above likelihood translates into an upper bound on $\sum m_{\nu}<2.16\,{\text{eV}}$ from terrestrial data.

We now briefly introduce the test statistics we use in order to quantify the existing tension between terrestrial and cosmological neutrino mass constraints (see Ref.~\cite{Gariazzo:2023joe} for more detailed discussions). The adopted tests are the parameter goodness-of-fit test (with test statistic $Q_{\text{DMAP}}$), the parameter differences test (with test statistic $\Delta$), and the Bayesian suspiciousness test (with suspiciousness parameter $p_S$). The $Q_{\text{DMAP}}$ test statistic evaluates the ``cost'' of explaining datasets together (i.e.\ with the same parameter values) as opposed to describing them separately (i.e.\ each dataset can chose its own preferred parameter values). Given two datasets $A$ and $B$, the test statistics is computed as:
\begin{equation}
\begin{split}
Q \equiv &-2\ln\mathcal{L}_{AB}(\hat\theta_{AB}) \\
&+ 2\ln\mathcal{L}_{A}(\hat\theta_{A}) + 2\ln\mathcal{L}_{B}(\hat\theta_{B}) \,,
\end{split}
\label{eq:PG}
\end{equation}
where $\hat\theta_D$ denotes the parameter values which ``best'' describe dataset $D$, and ${\cal L}$ denotes the likelihood for the datasets given the parameter values. In the context of Bayesian analyses, $\hat\theta_D$ is set to the ``maximum a posteriori'' parameter values (MAP, the point at which the posterior assumes its maximum value), which in general does depend on the prior choice: see Refs.~\cite{Raveri:2018wln,DES:2020hen}, where the corresponding test statistics is denoted by $Q_{\text{DMAP}}$ (difference of log-likelihoods at their MAP point).

The parameter differences test statistics instead measures the distance between posterior distributions for the parameters $\theta$ of two different datasets~\cite{Raveri:2019gdp,Raveri:2021wfz}. We define the difference as $\Delta\theta\equiv\theta_1-\theta_2$, where $\theta_1$ and $\theta_2$ are two points in the shared parameter space. If $A$ and $B$ are independent datasets, the posterior distribution for $\Delta\theta$ is given by the following:
\begin{eqnarray}
\mathcal{P}_\Delta(\Delta\theta)=\int\mathcal{P}_A(\theta)\mathcal{P}_B(\theta-\Delta\theta)\,d\theta \,.
\label{eq:PD}
\end{eqnarray}
The probability for a given parameter shift between the two posteriors is given by the following integral:
\begin{eqnarray}
\Delta= \int_{\mathcal{P}_\Delta(\Delta\theta)>\mathcal{P}_\Delta(0)}\mathcal{P}_\Delta(\Delta\theta) d\Delta\theta\, .
\label{eq:delta_par_shift}
\end{eqnarray}
Values of $\Delta$ close to $0$ indicate agreement between the two datasets, whereas values close to $1$ indicate a tension.

Finally, for what concerns the Bayesian suspiciousness, the starting point is the Bayesian evidence ratio, defined as follows:
\begin{eqnarray}
R\equiv \frac{\mathcal{Z}_{AB}}{\mathcal{Z}_{A}\mathcal{Z}_{B}} \,,
\label{eq:lnZratio}
\end{eqnarray}
where the numerator corresponds to the evidence when the datasets $A$ and $B$ are described by the same set of parameters $\theta$, whereas in the denominator different parameters may be preferred by the two datasets.~\footnote{Values of $R \gg 1$ $(\ll 1)$ would indicate agreement (disagreement) between the two datasets.} As discussed in Ref.~\cite{DES:2020hen}, $R$ depends on the prior volume in such a way that small values of $R$, indicative of a possible tension between datasets, can be artificially increased by increasing the prior volume. This is the reason why we do not directly use the Bayesian evidence ratio in what follows. We instead adopt the information ratio $I$, based on the Kullback-Leibler divergence, to remove the prior dependence. In particular, we start from the log-information ratio, given by:
\begin{eqnarray}
\ln I = \mathcal{D}_A+\mathcal{D}_B-\mathcal{D}_{AB}\,,
\label{eq:lnI}
\end{eqnarray}
where the Kullback-Leibler divergence is defined as:
\begin{eqnarray}
\mathcal{D}_D=\int \text{d}\theta \, \mathcal{P}_D\ln\left(\frac{\mathcal{P}_D}{\Pi}\right) \,.
\label{eq:KL_div}
\end{eqnarray}
Using the log-information ratio we can cancel the prior dependence of the Bayesian evidence ratio $R$ and define the suspiciousness parameter $S$ as follows~\cite{Handley:2019wlz}:
\begin{eqnarray}
\ln S\equiv\ln R-\ln I \,.
\label{eq:lnS}
\end{eqnarray}
Positive values of $\ln S$ indicate agreement among the datasets, while negative ones are indicative of tension. In what follows, we shall report the probability $p(S)\equiv p_S$. Under certain conditions, all of which are met in our cases, the tests statistics considered above (or functions thereof) follow $\chi^2$ distributions with an appropriate number of degrees of freedom (see Ref.~\cite{Gariazzo:2023joe} for full details), and we will report the significance of the resulting tensions (if any) in terms of equivalent number of standard deviations $N\sigma$, by converting probabilities into two-sided Gaussian standard deviations.

Aside from quantifying the tension between cosmological and terrestrial observations, it is also important to assess the level of internal consistency between the adopted cosmological probes. This allows us to determine if certain cosmological combinations are somewhat ``artificial'', in the sense that possible very tight upper limits on $\sum m_{\nu}$ may be driven by internal tensions between various cosmological datasets. To quantify the cosmology-internal tension (if any), we adopt the $Q_{\text{DMAP}}$ and suspiciousness metrics. We do not use the parameter differences metric given its dependence on the cosmological priors and parametrization (e.g.\ whether one uses $\theta_s$ or $H_0$ as a fundamental parameter). For the comparison between cosmology and terrestrial experiments, this dependence is not an issue, since there is no ambiguity as to the parameter of interest, i.e.\ $\sum m_{\nu}$, which is always well constrained. The cosmology-internal tension for various dataset combinations is calculated in Appendix~\ref{sec:appendixb}, see Tab.~\ref{tab:tension_between_datasets}: we anticipate that the only dataset combinations featuring a $>2\sigma$ internal tension, whose associated parameter constraints should therefore be considered somewhat artificial (as we will repeatedly stress in what follows), are those involving the SH0ES prior.

\section{Results}
\label{sec:results}

\begin{table*}[!t]
\centering
\begin{tabular}{|l?cc|cc|}
\hline \hline
& \multicolumn{2}{c|}{$\Lambda$CDM+$\sum m_{\nu}$}
& \multicolumn{2}{c|}{NPDDE+$\sum m_{\nu}$}\\
\cline{2-5}
\textbf{Dataset combination} & $\sum m_{\nu}\,[{\text{eV}}]$ & $B_{\text{NO,IO}}$ & $\sum m_{\nu}\,[{\text{eV}}]$ & $B_{\text{NO,IO}}$\\
\hline \hline
baseline (CMB + DESI) & $<0.072$ & 8.1 & $<0.064$ & $12.3$\\
baseline + SNeIa & $<0.081$ & 7.0 & $<0.068$ & $7.9$\\
baseline + CC & $<0.073$ & 7.3 & $<0.067$ & $8.0$ \\
baseline + SDSS & $<0.083$ & 6.8 & $<0.070$ & $10.6$ \\
baseline + SH0ES & $<0.048$ &47.8 & $<0.047$ & $54.6$ \\
baseline + XSZ & $<0.050$ &46.5 & $<0.044$ & $39.6$ \\
baseline + GRB & $<0.072$ & 8.7 & $<0.066$ & $15.4$ \\
\hline \hline
aggressive combination (baseline + SH0ES + XSZ) & $<0.042\,{\text{eV}}$ &72.6 & $<0.041\,{\text{eV}}$ & $109.2$ \\
\hline \hline
CMB (with ACT ``extended'' likelihood)+DESI & $<0.072$ & 8.0 & $<0.065$ & $12.8$ \\
CMB+DESI (with 2020 \texttt{HMCode})&$<0.074$ & 7.5 & $<0.065$ & $10.8$ \\
CMB (with \texttt{v1.2} ACT likelihood)+DESI & $<0.082$ & 7.4 & $<0.072$ & $6.3$ \\
CMB (with PR4 data)+DESI & $<0.080$ & 6.4 & $<0.064$ & $12.5$ \\
\hline \hline
\end{tabular}
\caption{$95\%$~C.L. upper limits on the sum of the neutrino masses $\sum m_{\nu}$ (in ${\text{eV}}$) and Bayes factor for normal ordering versus inverted ordering, $B_{\text{NO,IO}}$ (with values of $B_{\text{NO,IO}}>1$ indicating a preference for the normal ordering) in light of different dataset combinations as listed in the leftmost column, and within two different cosmological models: the 7-parameter $\Lambda$CDM+$\sum m_{\nu}$ model (two intermediate columns), and the 9-parameter NPDDE+$\sum m_{\nu}$ model where the dark energy equation of state is modeled as in Eq.~(\ref{eq:cpl}) and required to satisfy $w(z) \geq -1$ (two rightmost columns).}
\label{tab:bounds}
\end{table*}

\begin{figure}[!ht]
\centering
\includegraphics[width=0.9\linewidth]{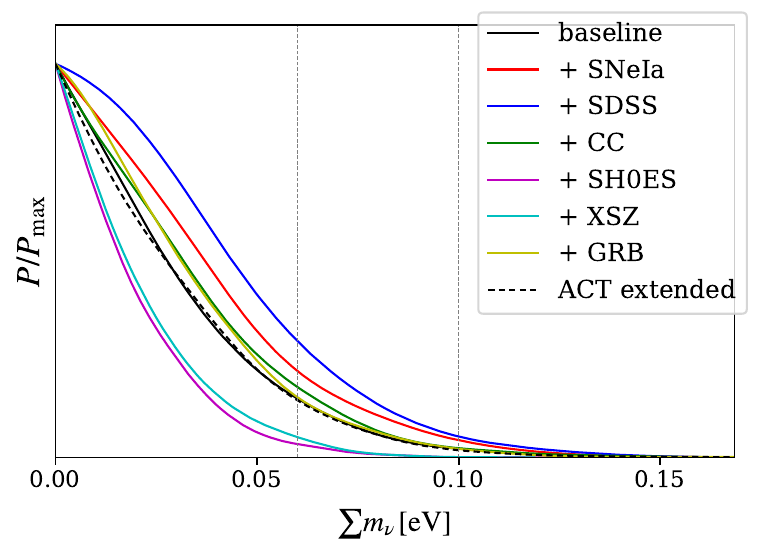}
\caption{Posterior distributions for the sum of the neutrino masses $\sum m_{\nu}$ (in ${\text{eV}}$) obtained within the 7-parameter $\Lambda$CDM+$\sum m_{\nu}$ model in light of different dataset combinations, as per the color coding.}
\label{fig:mnu_LCDM}
\end{figure}

\begin{figure}[!ht]
\centering
\includegraphics[width=0.9\linewidth]{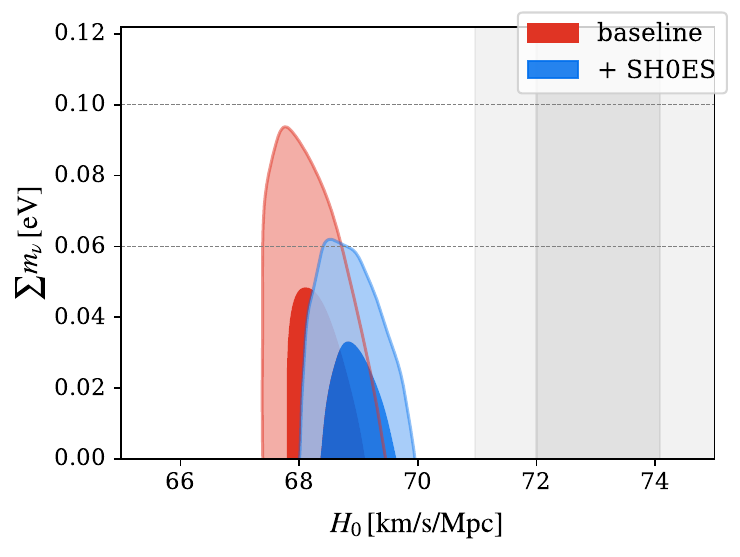}
\caption{2D joint posterior distribution for the sum of the neutrino masses $\sum m_{\nu}$ (in ${\text{eV}}$) and the Hubble constant $H_0$ (in ${\text{km}}/{\text{s}}/{\text{Mpc}}$) obtained within the 7-parameter $\Lambda$CDM+$\sum m_{\nu}$ model, and in light of the baseline dataset combination (red contours), and the combination of the latter with the SH0ES prior (blue contours). The (weak) anti-correlation between the two parameters explains why adding the SH0ES prior tightens the upper limits on $\sum m_{\nu}$. The grey band indicates the SH0ES measurement $H_0=(73.04 \pm 1.04)\,{\text{km}}/{\text{s}}/{\text{Mpc}}$ reported in Ref.~\cite{Riess:2021jrx}.}
\label{fig:mnu_H0_LCDM}
\end{figure}

We now discuss the limits we obtain on $\sum m_{\nu}$ from various combinations of the datasets presented previously, and quantify the preference for the NO versus the IO, as well as the tension between cosmological and terrestrial measurements. From now on, all upper limits on $\sum m_{\nu}$ are at 95\%~C.L. unless otherwise stated. A summary of our upper limits on $\sum m_{\nu}$ and the Bayes factors for the NO versus the IO when adopting various dataset combinations is provided in Tab.~\ref{tab:bounds}.

For the sake of comparison with previous works in the literature, we begin by reporting the results obtained within the $\Lambda$CDM+$\sum m_{\nu}$ model, analyzing the impact of likelihood settings. Posterior distributions for $\sum m_{\nu}$ in light of different dataset combinations are shown in Fig.~\ref{fig:mnu_LCDM}. For our baseline dataset combination of CMB data with the DESI BAO measurements, we find $\sum m_{\nu}<0.072\,{\text{eV}}$, in perfect agreement with the limit obtained by the DESI collaboration~\cite{DESI:2024mwx}. In this case, the code used to treat non-linearities is the 2016 version of \texttt{HMCode}, whereas we use the \texttt{v1.1} version of the ACT lensing likelihood, in agreement with the settings adopted by DESI. If we instead use the same version of \texttt{HMCode} but switch to the \texttt{v1.2} version of the ACT lensing likelihood, the previous limit relaxes to $\sum m_{\nu}<0.082\,{\text{eV}}$.~\footnote{The former limit further degrades to $\sum m_\nu<0.084\,{\text{eV}}$ when the effects of baryonic feedback is included in the data analyses.} Using the ``extended'' version of the ACT likelihood instead has virtually no effects on the derived limit. Finally, if the \texttt{v1.1} version of the ACT lensing likelihood is used together with the 2020 version of \texttt{HMCode}, the limit we obtain is $\sum m_{\nu}<0.074\,{\text{eV}}$. Therefore, assumptions concerning the effect of non-linearities lead to very mild shifts in the obtained limits. On the other hand, CMB lensing assumptions have a more important effect, given the central role played by CMB lensing data when obtaining limits on $\sum m_{\nu}$ from CMB-only data. Finally, one may wonder how stable our constraints are against the assumed Big Bang Nucleosynthesis model. Replacing the default \texttt{PRIMAT} model with the \texttt{ParthENoPE} one, while using the 2016 version of \texttt{HMCode} and the \texttt{v1.2} version of the ACT likelihood changes the limit from $\sum m_{\nu}<0.082\,{\text{eV}}$ to $\sum m_{\nu}<0.081\,{\text{eV}}$, showing that the choice of BBN model plays a negligible role.

Our baseline upper limit of $\sum m_{\nu}<0.072\,{\text{eV}}$ is also stable against the inclusion of additional datasets -- see Fig.~\ref{fig:mnu_LCDM} for $\sum m_{\nu}$ posteriors within the $\Lambda$CDM+$\sum m_{\nu}$ model in light of various dataset combinations, or Tab.~\ref{tab:bounds} for the numerical results. For instance, we see that adding CC measurements barely changes the bound, which is now $\sum m_{\nu}<0.073\,{\text{eV}}$. On the other hand, the inclusion of SDSS data with the procedure discussed in Sec.~\ref{subsec:data}, as well as in Sec.~3.3 and Appendix~A of Ref.~\cite{DESI:2024mwx}, slightly degrades the bound to $\sum m_{\nu}<0.083\,{\text{eV}}$. A very similar limit of $\sum m_{\nu}<0.081\,{\text{eV}}$ is obtained when the \textit{PantheonPlus} SNeIa dataset is combined with our baseline dataset.

The tightest limit is obtained when combining our baseline dataset with the SH0ES prior, due to the well-known anti-correlation between $\sum m_{\nu}$ and $H_0$. Indeed, once $H_0$ is raised, $\sum m_{\nu}$ (and more generally $\Omega_m$) needs to be lowered in order to keep $\theta_s$ fixed, as extensively discussed for instance in Refs.~\cite{DiValentino:2015sam,Giusarma:2016phn,Vagnozzi:2017ovm}. In this case, we obtain the extremely tight limit $\sum m_{\nu}<0.048\,{\text{eV}}$, although we caution that this limit cannot be deemed reliable due to the Hubble tension (see e.g.\ Refs.~\cite{Verde:2019ivm,DiValentino:2021izs,Perivolaropoulos:2021jda,Schoneberg:2021qvd,Shah:2021onj,Abdalla:2022yfr,DiValentino:2022fjm,Hu:2023jqc,Vagnozzi:2023nrq,Verde:2023lmm} for reviews), which is formally exacerbated when introducing a non-zero $\sum m_{\nu}$ (increasing $H_0$ at fixed $\theta_s$ would require $\sum m_{\nu}<0\,{\text{eV}}$). The results obtained using the SH0ES prior serve as a warning of the danger of combining inconsistent datasets, especially within the context of models which would formally worsen the Hubble tension (see a similar discussion in this context in Ref.~\cite{Vagnozzi:2017ovm}). Therefore, limits obtained using the SH0ES prior are to be considered artificial. We recall that the level of internal tension between cosmological probes is quantified in Appendix~\ref{sec:appendixb}, see Tab.~\ref{tab:tension_between_datasets}. There we see that the tension between our baseline dataset combination and the SH0ES prior always exceeds $3\sigma$, independently of the underlying model ($\Lambda$CDM+$\sum m_{\nu}$ and NPDDE+$\sum m_{\nu}$) and tension metric ($Q_{\text{DMAP}}$ and suspiciousness). This confirms that any result obtained using the SH0ES prior should be considered artificial, and driven by the cosmology-internal tension. As such, these results should be taken with a significant grain of salt.

Our baseline limit of $0.072\,{\text{eV}}$ can be further improved if one considers the XSZ and GRB datasets, which improve the determination of the background expansion rate and are therefore crucial in further breaking the geometrical degeneracy. Adding the XSZ dataset to our baseline combination significantly improves the limit to $0.050\,{\text{eV}}$, close to the artificial limit obtained when combining with the SH0ES prior on the Hubble constant. However, we stress that this combination is not artificial, since the XSZ dataset is always in better than $2\sigma$ agreement with the baseline dataset combination, regardless of underlying model and tension metric. This is also clear from Fig.~\ref{fig:distance}, from which one sees that the XSZ band is in good agreement with the baseline dataset combination, while lying in a region of parameter space which can naturally be accommodated by lower values of $\sum m_{\nu}$. On the other hand, combining our baseline dataset with the GRB dataset does not change our limit, which remains $0.072\,{\text{eV}}$. This is most likely due to the well-known and documented preference for larger values of $\Omega_m$, which can naturally be accommodated with larger values of $\sum m_{\nu}$ given that massive neutrinos contribute to the density of non-relativistic matter at late times (see Fig.~\ref{fig:distance}, where the fact that GRB should lead to a preference for larger values of $\sum m_{\nu}$ is evident). Finally, purely for illustrative purposes, we consider an ``aggressive'' dataset combination where the SH0ES and XSZ datasets are added to our baseline dataset. In this case we find that our upper limit improves to $0.042\,{\text{eV}}$, with all the caveats discussed previously. We see that within all three the baseline+SH0ES, baseline+XSZ, and aggressive dataset combinations, the 95\%~C.L. upper limit on $\sum m_{\nu}$ is clearly in tension with the lower limit $\sum m_{\nu}>0.06\,{\text{eV}}$ arising from oscillation experiments, indicating a tension between cosmological and terrestrial observations which will be quantified later. However, we also note that the first and third combinations are ``artificial'', in the sense that the resulting tight constraints are driven by internal tensions (see Appendix~\ref{sec:appendixb}, Tab.~\ref{tab:tension_between_datasets}). On the other hand, this problem is not present for the XSZ dataset: the resulting constraints are therefore deemed reliable.

\begin{figure}[!ht]
\centering
\includegraphics[width=0.9\linewidth]{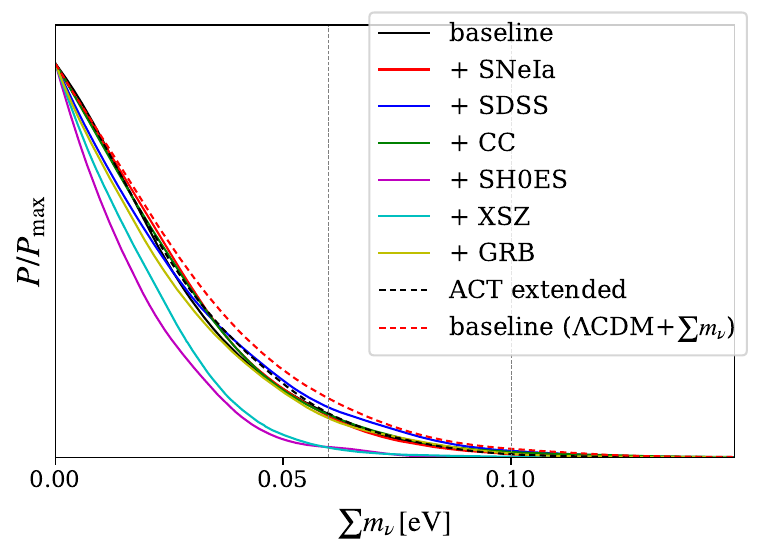}
\caption{As in Fig.~\ref{fig:mnu_LCDM}, but within the 9-parameter NPDDE+$\sum m_{\nu}$ model. For comparison we also include the posterior obtained from the baseline dataset combination within the 7-parameter $\Lambda$CDM+$\sum m_{\nu}$ model (red dashed curve, which should be directly compared against the solid black curve).}
\label{fig:mnu_wzgeq-1}
\end{figure}

\begin{figure}[!ht]
\centering
\includegraphics[width=\linewidth]{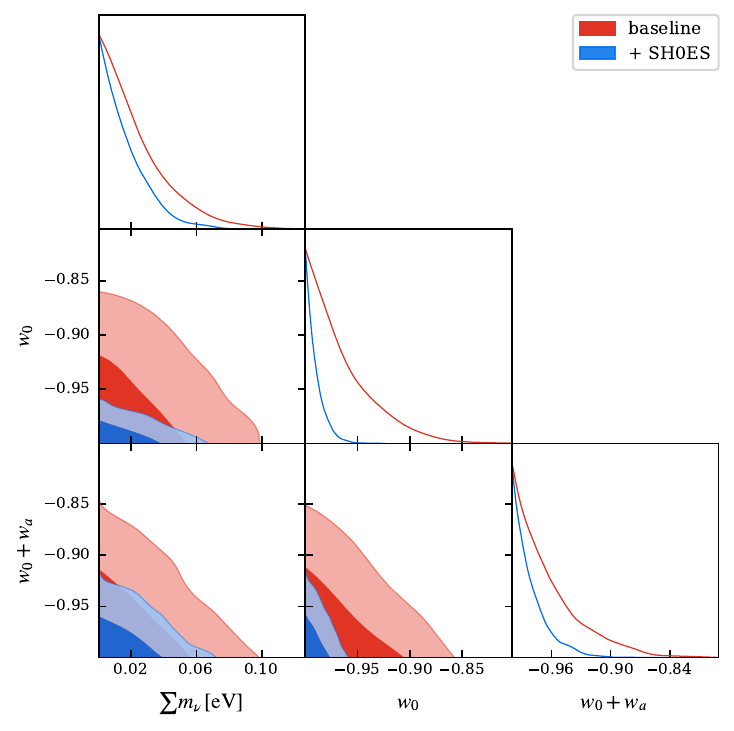}
\caption{Triangular plot showing 2D joint and 1D marginalized posterior probability distributions for the sum of the neutrino masses $\sum m_{\nu}$ (in ${\text{eV}})$, the present-day dark energy equation of state $w_0$, and the parameter combination $w_0+w_a$ (which is required to be $\geq -1$ within the NPDDE region of parameter space), obtained within the 9-parameter NPDDE+$\sum m_{\nu}$ model in light of the baseline dataset combination (red curves), and the combination of the latter with the SH0ES prior (blue contours). }
\label{fig:mnu_w_wa_wzgeq-1}
\end{figure}

\begin{figure}[!ht]
\centering
\includegraphics[width=\linewidth]{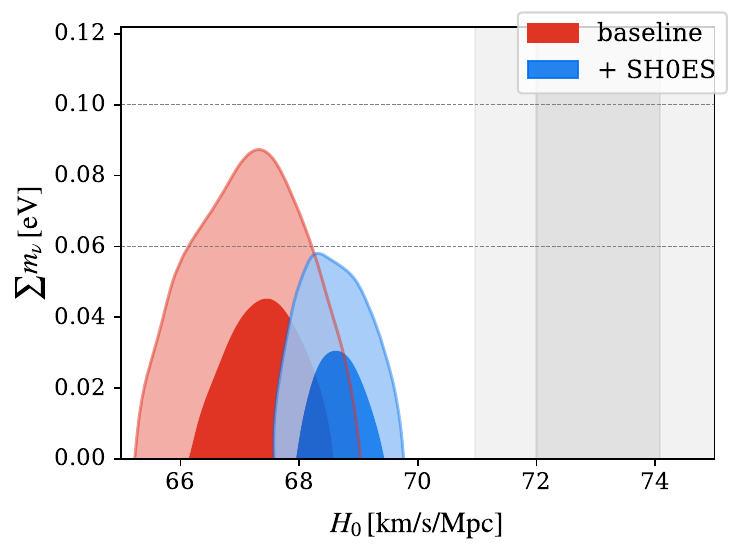}
\caption{As in Fig.~\ref{fig:mnu_H0_LCDM}, but within the 9-parameter NPDDE+$\sum m_{\nu}$ model, showing that the degeneracy between $\sum m_{\nu}$ and $H_0$ is only weakly affected by allowing for a time-varying dark energy component. The grey band indicates the SH0ES measurement $H_0=(73.04 \pm 1.04)\,{\text{km}}/{\text{s}}/{\text{Mpc}}$~\cite{Riess:2021jrx}.}
\label{fig:mnu_H0_wzgeq-1}
\end{figure}

As discussed in Sec.~\ref{subsec:models}, we now move on to consider the physically motivated NPDDE+$\sum m_{\nu}$ model, where the evolving DE component is required to lie in the quintessence-like $w(z) \geq -1$ region in order for the null energy condition not to be violated (the case where $w(z)$ can also enter the phantom region is briefly discussed in Appendix~\ref{sec:appendix}). Posterior distributions for $\sum m_{\nu}$ in light of different dataset combinations are shown in Fig.~\ref{fig:mnu_wzgeq-1}. Compared to $\Lambda$CDM, and at fixed $H_0$, a DE component with $w<-1$ ($w>-1$) pushes the CMB further away from (closer to) us, and therefore needs to be compensated by larger (smaller) values of $\sum m_{\nu}$ in order to keep $\theta_s$ fixed. This can be clearly seen in the triangular plot of Fig.~\ref{fig:mnu_w_wa_wzgeq-1}. As shown in Ref.~\cite{Vagnozzi:2018jhn}, restricting to the $w(z) \geq -1$ region results in the upper limits on $\sum m_{\nu}$ actually tightening in spite of the extended parameter space (see also footnote~4) and the discussion in Ref.~\cite{Green:2024xbb} for further clarifications). In this case, we see that allowing for a time-varying $w(z)$ does not substantially alter the degeneracy between $H_0$ and $\sum m_{\nu}$: see Fig.~\ref{fig:mnu_H0_wzgeq-1} and compare it to Fig.~\ref{fig:mnu_H0_LCDM}. The upper limit we find on $\sum m_{\nu}$ for our baseline dataset combination is $0.064\,{\text{eV}}$, already in mild tension with the lower limit from oscillation experiments. This is tighter than the limit within the $\Lambda$CDM+$\sum m_{\nu}$ model, confirming once more the earlier findings reported in Ref.~\cite{Vagnozzi:2018jhn}. As within the $\Lambda$CDM+$\sum m_{\nu}$ model, we see a very similar trend when changing analyses settings pertaining to the treatment of non-linearities, use of the ``extended'' ACT lensing likelihood, and use of the \texttt{v1.1} versus \texttt{v1.2} version of the ACT likelihood, all of which have at best mild effects on the derived limits (whereas once more the choice of BBN models has no effect).

Again, the inclusion of SNeIa, CC, SDSS BAO, or GRB data has a very marginal effect on the previous limit (refer to the right-most columns of Tab.~\ref{tab:bounds}). The most notable improvements are once again observed when including the SH0ES prior (again with the same warnings and caveats as before), in which case the limit improves to $\sum m_{\nu}<0.047\,{\text{eV}}$, whereas the inclusion of the XSZ datasets tightens the bound to $\sum m_{\nu}<0.044\,{\text{eV}}$. Finally, in our most aggressive dataset combination we find the extremely tight upper limit $\sum m_{\nu}<0.041\,{\text{eV}}$.

It is clear even just at a qualitative level from all the limits discussed so far, and reported in Tab.~\ref{tab:bounds} (where we also report $B_{\rm NO,IO}$), that the NO will be preferred over the IO (see also Ref.~\cite{Du:2024pai}). Within the $\Lambda$CDM+$\sum m_{\nu}$ model, the preference for the NO versus the IO is always at least \textit{positive}. This is the case for all dataset combinations which do not include the SH0ES and/or XSZ datasets, which lead to the tightest limits. Once the SH0ES and/or XSZ datasets are added to our baseline combination, the preference for the NO versus the IO is instead always \textit{strong} (i.e.\ $3\leq\ln B_{\rm NO,IO}<5$), a fact which remains true even in our most aggressive dataset combination. Very similar features are observed in the NPDDE+$\sum m_{\nu}$ model, albeit typically with slightly stronger preferences for the NO versus the IO.

\begin{figure*}[!ht]
\centering
\includegraphics[width=0.8\linewidth]{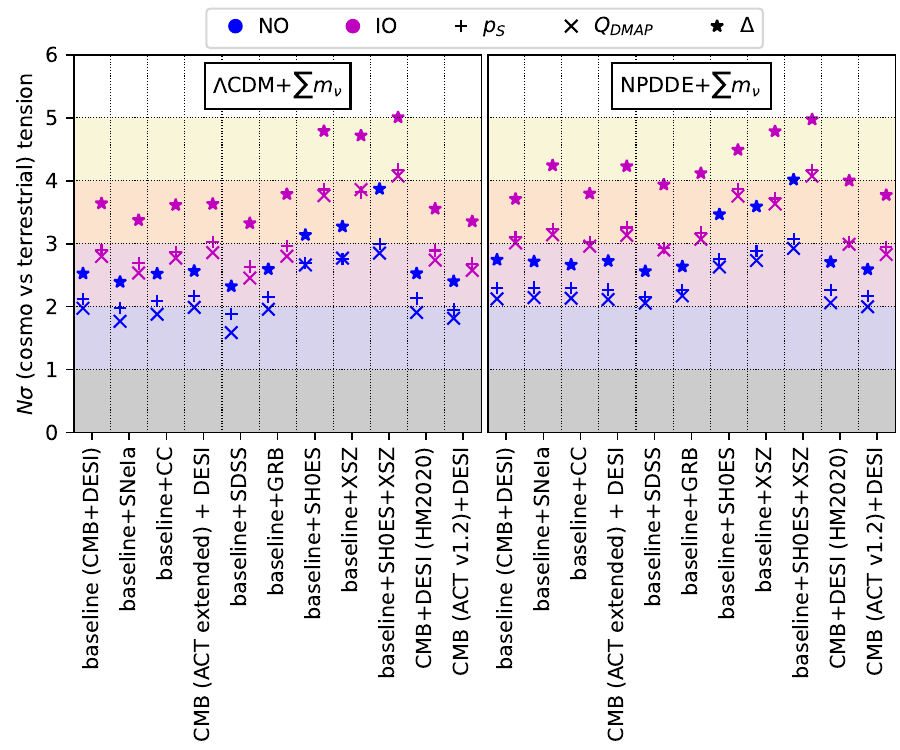}
\caption{Tension between cosmological and terrestrial observations in equivalent number of standard deviations, in light of different dataset combinations as indicated on the abscissa axis, obtained within the 7-parameter $\Lambda$CDM+$\sum m_{\nu}$ (left panel) and 9-parameter NPDDE+$\sum m_{\nu}$ (right panel) cosmological models, for both the normal ordering (blue markers) and inverted ordering (magenta markers). The tension has been computed for the three different test statistics discussed: $Q_{\text{DMAP}}$ (crosses), parameter differences $\Delta$ (stars), and Bayesian suspiciousness $p_S$ (pluses). Broadly speaking, we see a $\approx 3\sigma$ tension between cosmology and terrestrial experiments.}
\label{fig:tensions}
\end{figure*}

More interesting is the quantification of the tension between cosmology and terrestrial experiments, as it is the first time that this is being done in light of data which clearly displays such a tension, given that most of the bounds reported previously disagree with the $\sum m_{\nu}>0.06\,{\text{eV}}$ lower limit. The results of our analysis are shown in Fig.~\ref{fig:tensions}, where we display the tension in equivalent number of standard deviations and given different dataset combinations for the $\Lambda$CDM+$\sum m_{\nu}$ (left panel) and NPDDE+$\sum m_{\nu}$ (right panel) models, for both the NO (blue markers) and IO (magenta markers) and for the three different test statistics discussed: $Q_{\text{DMAP}}$ (crosses), parameter differences $\Delta$ (stars) and Bayesian suspiciousness $p_S$ (pluses). From the same Figure, we notice that the highest (lowest) significance is always achieved when adopting the $\Delta$ ($Q_{\text{DMAP}}$) test statistic, with differences typically of order $0.5\sigma$, but potentially as large as $1\sigma$, between the two (whereas in general we find good agreement between $Q_{\text{DMAP}}$ and Bayesian suspiciousness). For all models, orderings, dataset combinations and test statistics adopted, however, the level of tension always falls between $2\sigma$ and $5\sigma$.

More in detail, focusing on the $\Delta$ test statistic for concreteness, we find that for all dataset combinations which do not include either SH0ES or XSZ, the level of tension between cosmological and terrestrial experiments for the NO is around $2.5\sigma$ when DE is in the form of a cosmological constant, and closer to $3\sigma$ when considering a quintessence-like evolving DE component. For the IO, the previous figures increase to the $\approx 3.5\sigma$ and $\approx 4\sigma$ levels respectively. Once either or both the SH0ES or XSZ datasets are included, the tension reaches the $3.5\sigma$ level for the NO for both cosmological models, while it approaches or even surpasses the $5\sigma$ level for the IO in both models. Even when considering the most conservative case overall, we can conclude that there is a $2.5\sigma$ tension within the NO, and a $3.5\sigma$ tension within the IO. Broadly speaking, we therefore conclude that there is a $\approx 3\sigma$ tension between cosmology and terrestrial experiments, which is evident in light of the very tight upper limits on $\sum m_{\nu}$ reported.

\section{Conclusions}
\label{sec:conclusions}

While officially ushering us into the era of Stage IV cosmology, the DESI BAO measurements have also led to a number of puzzles. Aside from issues related to the nature of dark energy, the extremely tight upper limits on the sum of the neutrino masses $\sum m_{\nu}$ (uncomfortably close to scratching the surface of the minimum value allowed by terrestrial oscillations $\sum m_{\nu} \gtrsim 0.06\,{\text{eV}}$) also raise several important questions. For instance, one may ask whether the limit reported by the DESI collaboration $\sum m_{\nu}<0.072\,{\text{eV}}$~\cite{DESI:2024mwx} can be further improved by considering other available cosmological probes, perhaps beyond the most widely used ones, and to what extent these limits result in the normal ordering (NO) being preferred against the inverted ordering (IO). In addition, tight upper limits on $\sum m_{\nu}$ are indicative of potential tensions between cosmological and terrestrial observations, whose proper quantification is important. Our work aims to address these questions (see also the very recent Ref.~\cite{Du:2024pai} for a related study, whose results agree with ours).

Considering a wide range of available late-time background probes, we further strengthen the upper limit reported by the DESI collaboration~\cite{DESI:2024mwx} down to $\sum m_{\nu}<0.050\,{\text{eV}}$ within a 7-parameter $\Lambda$CDM+$\sum m_{\nu}$ model \textit{even when we do not consider the SH0ES prior on the Hubble constant}. Adding the latter results in our tightest upper limit being $\sum m_{\nu}<0.042\,{\text{eV}}$, although this limit is somewhat artificial as driven by the Hubble tension. In all dataset combinations we have explored, the level of preference for the NO versus the IO is always substantial or strong, with the Bayes factor for NO against IO being as large as $46.5$ when not considering the SH0ES $H_0$ prior. We have also analyzed the impact of various analysis assumptions/likelihood settings (e.g.\ use of ``extended'' ACT lensing likelihood and version thereof, treatment of non-linearities, and BBN model), finding all of these to have a very minor effect. We have then robustly quantified the level of tension between cosmological and terrestrial observations. This is found to depend on the specific test statistic adopted ($Q_{\text{DMAP}}$, parameter differences $\Delta$, or suspiciousness, see Fig.~\ref{fig:tensions}), but is always between $2.5\sigma$ and $5\sigma$ for the NO, and between $3\sigma$ and $5\sigma$ for the IO. Our conservative conclusion is therefore that there is broadly a $\approx 3\sigma$ tension between cosmology and terrestrial experiments, evident from the very tight upper limits on $\sum m_{\nu}$. Finally, we have extended all these results to the case where the dark energy component is evolving but restricted to lie in the physically motivated non-phantom regime $w(z) \geq -1$ (as expected for instance in the simplest quintessence models), finding that all of the previous figures are (very slightly) tightened, in agreement with earlier results~\cite{Vagnozzi:2018jhn}. This highlights once more a potentially interesting synergy between the nature of dark energy and laboratory determinations of the neutrino mass ordering (e.g.\ via long-baseline experiments).

The above results have potentially very far-reaching consequences for what concerns the search for new physics, both on the cosmology and particle physics sides. Besides the obvious implications for the neutrino mass ordering and for neutrino model-building, the tension between cosmology and terrestrial experiments which we have quantified to be potentially as large as $5\sigma$ calls for an explanation, which could require new cosmological physics (e.g.\ a primordial trispectrum resembling the signal from CMB lensing~\cite{Green:2024xbb}, models of dark energy and/or modified gravity which could substantially weaken cosmological limits on $\sum m_{\nu}$~\cite{Elbers:2024sha} or even ``hide'' them~\cite{Bellomo:2016xhl}, and so on) or new particle physics (e.g.\ time-varying neutrino masses from cosmic phase transitions~\cite{Lorenz:2018fzb}, new long-range forces~\cite{Green:2024xbb}, interactions with dark matter~\cite{Sen:2024pgb}, a non-standard distribution~\cite{Oldengott:2019lke,Alvey:2021sji}, or more generally non-standard neutrino physics~\cite{Beacom:2004yd,Farzan:2015pca,Chacko:2019nej,Chacko:2020hmh,Escudero:2020ped,Esteban:2021ozz,Escudero:2022gez}), in any event with dramatic consequences for fundamental physics. However, we choose to remain cautiously optimistic with our claims, given a number of caveats surrounding current neutrino mass cosmological constraints -- for instance, the alleged preference for negative $\sum m_{\nu}$, which we have here chosen not to explore by imposing the physically-motivated positivity of $\sum m_{\nu}$ (although we recognize the value of studying the $\sum m_{\nu}<0$ region as a consistency test), and more generally the impact of the lensing anomaly on $\sum m_{\nu}$ constraints~\cite{DiValentino:2021imh,DiValentino:2023fei,Giare:2023aix}. It is reasonable to expect that these issues will be clarified with upcoming CMB data~\cite{CMB-S4:2016ple,SimonsObservatory:2018koc,SimonsObservatory:2019qwx} as well as improved analyses of current CMB data (see Appendix~\ref{sec:appendixc}), and the release of more data from DESI (we note that the full-shape results have recently been released~\cite{DESI:2024jis,DESI:2024yrg,DESI:2024hhd}), as well as data from other large-scale structure surveys such as \textit{Euclid}~\cite{Amendola:2016saw}. In a few years, we will therefore have either a convincing detection of non-zero $\sum m_{\nu}$, or a convincing non-detection thereof, with a consequential strong tension between cosmology and terrestrial observations of which we are perhaps catching the first glimmer: regardless of the outcome either of these scenarios will be game-changing, and will lead to neutrinos having realized in full their promise as portals onto new physics.

\begin{acknowledgments}
\noindent J.-Q.J.\ acknowledges support from the Joint PhD Training program of the University of Chinese Academy of Sciences. W.G.\ acknowledges support from the Lancaster–Sheffield Consortium for Fundamental Physics through the Science and Technology Facilities Council (STFC) grant ST/X000621/1. S.G.\ acknowledges support from the La Caixa Foundation and the European Union Framework Programme for Research and Innovation Horizon 2020 (2014–2020) through a La Caixa Junior Leader Fellowship LCF/BQ/PI23/11970034, and from the Istituto Nazionale di Fisica Nucleare (INFN) through the Commissione Scientifica Nazionale 4 (CSN4) Iniziativa Specifica ``Theoretical Astroparticle Physics'' (TAsP). M.G.D.\ acknowledges support from the Division of Science of the National Astronomical Observatory of Japan. E.D.V.\ acknowledges support from the Royal Society through a Royal Society Dorothy Hodgkin Research Fellowship. O.M.\ acknowledges support from the Ministerio de Ciencia, Innovaci\'{o}n y Universidades (MCIU, Spanish Ministry of Science, Innovation and Universities) with funding from the European Union NextGenerationEU (PRTR-C17.I01) and the Generalitat Valenciana (ASFAE/2022/020), and partial support by the Spanish MCIN/AEI/10.13039/501100011033 grants PID2020-113644GB-I00, by the European Innovative Training Networks (ITN) project HIDDeN (H2020-MSCA-ITN-2019/860881-HIDDeN) and Staff Exchanges (SE) project ASYMMETRY (HORIZON-MSCA-2021-SE-01/101086085-ASYMMETRY), as well as by the Generalitat Valenciana grants PROMETEO/2019/083 and CIPROM/2022/69. O.M.\ acknowledges the Galileo Galilei Institute for Theoretical Physics (GGI) for hospitality during the completion of this work. D.P., S.S.C., and S.V.\ acknowledge support from the INFN CSN4 Iniziativa Specifica ``Quantum Fields in Gravity, Cosmology and Black Holes'' (FLAG). S.S.C.\ acknowledges support from the Fondazione Cassa di Risparmio di Trento e Rovereto (CARITRO Foundation) through a Caritro Fellowship (project ``Inflation and dark sector physics in light of next-generation cosmological surveys''). S.V. acknowledges support from the University of Trento and the Provincia Autonoma di Trento (PAT, Autonomous Province of Trento) through the UniTrento Internal Call for Research 2023 grant ``Searching for Dark Energy off the beaten track'' (DARKTRACK, grant agreement no.\ E63C22000500003). This publication is based upon work from the COST Action CA21136 ``Addressing observational tensions in cosmology with systematics and fundamental physics'' (CosmoVerse), supported by COST (European Cooperation in Science and Technology). We acknowledge the use of high performance computing services provided by the International Centre for Theoretical Physics Asia-Pacific cluster, TianHe-2 supercomputer and the National Observatory of Rio de Janeiro Data Center (DCON).
\end{acknowledgments}

\appendix

\section{Including the phantom region}
\label{sec:appendix}

\begin{figure}[!ht]
\centering
\includegraphics[width=\linewidth]{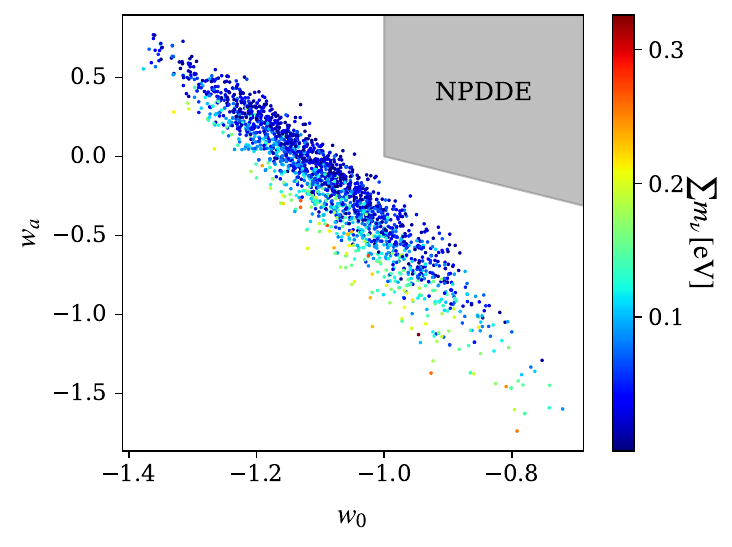}
\caption{2D scatter plot of MCMC samples in the $w_0$-$w_a$ plane, colored by the value of the sum of the neutrino masses $\sum m_{\nu}$ (in ${\text{eV}}$), obtained within the 9-parameter $w_0w_a$CDM model in light of the most aggressive dataset combination adopted (baseline+SH0ES+XSZ). The shaded part indicates the NPDDE region of parameter space.}
\label{fig:phantom}
\end{figure}

For completeness, we examine the impact of allowing our DE component to enter the phantom region, $w(z)<-1$. This is particularly relevant since the DESI BAO data, when interpreted within the $w_0w_a$CDM model, appear to prefer $w_a<0$, and therefore models which cross the phantom divide in the past~\cite{DESI:2024uvr,DESI:2024lzq}. We therefore check how our main results change if we consider the 9-parameter $w_0w_a$CDM+$\sum m_{\nu}$ model, where the DE EoS $w(z)$ is modeled following Eq.~(\ref{eq:cpl}), without the NPDDE constraint. In Fig.~\ref{fig:phantom} for illustrative purposes we show a 2D scatter plot of MCMC samples in the $w_0$-$w_a$ plane colored by the associated values of $\sum m_{\nu}$, in light of the most aggressive dataset combination adopted (baseline+SH0ES+XSZ). We note the very strong degeneracy between the two DE parameters, and that higher (lower) values of $\sum m_{\nu}$ are associated to regions of parameter space where $w_0<-1$ ($w_0>-1$) as expected. From Fig.~\ref{fig:phantom} one observes two things: \textit{a)} the posterior is completely outside the non-phantom region once one allows for generic values of $w_0$ and $w_a$, and \textit{b)} the non-phantom region can only be accessed if one allows negative values of $\sum m_{\nu}$. Within this model, we find the 95\%~C.L. upper limit $\sum m_{\nu}<0.159\,{\text{eV}}$, a factor of $\gtrsim 3$ weaker than the limit obtained within the $\Lambda$CDM+$\sum m_{\nu}$ model from the same dataset combination. Within the same setting we do not find a significant preference for the normal ordering, with $B_{\text{NO,IO}}=2.1$, whereas as expected there is no significant tension between cosmology and terrestrial observations.

\section{Internal tension between cosmological datasets}
\label{sec:appendixb}

As stressed several times throughout the manuscript, it is important to assess the level of internal agreement (or tension) between cosmological probes. We do so adopting the $Q_{\text{DMAP}}$ and suspiciousness metrics introduced earlier, which we use to compute the level of tension between the baseline (CMB+DESI) dataset combination and the other external probes we add. However, we do not compute the level of tension between baseline and SDSS, due to the fact that some of the DESI datapoints are removed when combining with SDSS (in any case, the level of tension between DESI and SDSS has been discussed in detail in other papers~\cite{DESI:2024mwx}). The results of our analysis are reported in Tab.~\ref{tab:tension_between_datasets}, where we see that the level of internal agreement (or tension) is always better than $2\sigma$ for all external datasets, except for the SH0ES prior (this is true regardless of model and tension metric), in which case there is always a $>3\sigma$ tension. Therefore, as stressed throughout the paper, any result obtained using the SH0ES prior should be considered artificial, and driven by this internal tension. However, the XSZ dataset is in $<2\sigma$ agreement with the baseline dataset combination (as is clear from Fig.~\ref{fig:distance}), and therefore results obtained using this dataset, while pushing towards tighter constraints on $\sum m_{\nu}$, can be considered trustworthy, or at the very least not driven by cosmology-internal tension.

\begin{table}[!ht]
\centering
\begin{tabular}{|l?cc|cc|}
\hline \hline
& \multicolumn{2}{c|}{$\Lambda$CDM+$\sum m_{\nu}$}
& \multicolumn{2}{c|}{NPDDE+$\sum m_{\nu}$}\\
\cline{2-5}
Dataset & $Q_\text{DMAP}$ & $S$ & $Q_\text{DMAP}$ & $S$ \\ \hline
SNeIa & $0.8\sigma$ & $1.1\sigma$ & $0.3\sigma$ & $0.7\sigma$ \\
CC & $0.5\sigma$ & $0.5\sigma$ & $<0.1\sigma$ & $0.4\sigma$ \\
SH0ES & $4.0\sigma$ & $3.2\sigma$ & $3.5\sigma$ & $3.1\sigma$ \\
XSZ & $1.9\sigma$ & $1.4\sigma$ & $1.1\sigma$ & $1.3\sigma$\\
GRB & $0.8\sigma$ & $0.9\sigma$ & $0.6\sigma$ & $0.7\sigma$\\
\hline \hline
\end{tabular}
\caption{Tension between our baseline (CMB+DESI) dataset combination and other external probes, for the two models and tension metrics considered.}
\label{tab:tension_between_datasets}
\end{table}

\section{Impact of Planck PR4 likelihoods}
\label{sec:appendixc}

\begin{table*}[!ht]
\centering
\begin{tabular}{|l?cc|cc|}
\hline \hline
& \multicolumn{2}{c|}{$\Lambda$CDM+$\sum m_{\nu}$}
& \multicolumn{2}{c|}{NPDDE+$\sum m_{\nu}$}\\
\cline{2-5}
\textbf{Dataset combination} & $\sum m_{\nu}\,[{\text{eV}}]$ & $B_{\text{NO,IO}}$ & $\sum m_{\nu}\,[{\text{eV}}]$ & $B_{\text{NO,IO}}$\\
\hline
CMB (PR4)+DESI & $<0.080$ & 6.4 & $<0.064$ & 12.5 \\
CMB (PR4)+DESI+SNeIa & $<0.090$ & 6.4 & $<0.070$ & 11.4 \\
CMB (PR4)+DESI+SDSS & $<0.090$ & 5.7 & $<0.078$ & 6.0 \\
\hline \hline
\end{tabular}
\caption{Impact on the resulting $\sum m_{\nu}$ constraints of the use of the \textit{Planck} \texttt{HiLLiPoP} and \texttt{LoLLiPoP} PR4 likelihoods, instead of their PR3 counterparts, tested on three dataset combinations.}
\label{tab:PR4}
\end{table*}

To examine the impact of the choice of \textit{Planck} likelihoods, we consider three dataset combinations and replace part of the \textit{Planck} PR3 likelihoods with their updated PR4 counterparts. Specifically, we replace the high-$\ell$ ($\ell>30$) part with the \texttt{HiLLiPoP} likelihood~\cite{Tristram:2023haj} for high-$\ell$ TT, TE, and EE spectra, and the low-$\ell$ ($\ell \leq 30$) part with the \texttt{LoLLiPoP} likelihood~\cite{Hamimeche:2008ai,Mangilli:2015xya,Tristram:2020wbi} for low-$\ell$ EE spectra. Such likelihoods have been shown to lead to a value of $A_L = 1.039 \pm 0.052$ for the lensing amplitude~\cite{Tristram:2023haj}, in much better agreement with the value expected within $\Lambda$CDM: we can thus expect the constraints on $\sum m_{\nu}$ to slightly degrade when switching to the PR4 likelihoods, given that larger values of $\sum m_{\nu}$ lead to less structure, and therefore to less lensing experienced by CMB photons on their way to us. On the other hand, we still adopt the PR4 lensing likelihood, as well as the \texttt{Commander} low-$\ell$ TT likelihood.

The results of the analysis are shown in Tab.~\ref{tab:PR4}, for three specific combinations we choose as case study: CMB+DESI, CMB+DESI+SNeIa, and CMB+DESI+SDSS, where in all three cases the CMB dataset includes the PR4 likelihoods as discussed above. We find that in all three cases the upper limits on $\sum m_{\nu}$ degrade slightly, by approximately $0.008\,{\text{eV}}$, in excellent agreement with the findings of Ref.~\cite{Allali:2024aiv}. The Bayes factors for NO versus IO also get correspondingly weaker. On the other hand, the tightening of the constraints as we move from the $\Lambda$CDM+$\sum m_{\nu}$ model to the NPDDE+$\sum m_{\nu}$ model persists. This is completely in line with our expectations, as the tightening is purely a prior effect~\cite{Vagnozzi:2018jhn}, and is therefore not expected to be affected by the adopted likelihoods. Although we have only considered three case studies, there is no reason to believe that these findings should not hold with other dataset combinations. We conclude that our main results are stable against the use of the \textit{Planck} PR4 likelihoods in place of the PR3 ones.

\section{SH0ES prior: $H_0$ versus $M_B$}
\label{sec:appendixd}

It has been argued in Refs.~\cite{Camarena:2021jlr,Efstathiou:2021ocp} that the proper way of including the SH0ES information is not through a prior on $H_0$, but rather on the SNeIa absolute magnitude $M_B$, since the SH0ES determination of $H_0$ is not directly done at $z=0$ but extrapolating the calibrated distance-redshift diagram of the SNeIa calibrator sample. However, the difference between the two approaches is expected to be substantial only for very non-smooth models of the late-time expansion history, particularly those where the expansion rate changes abruptly at extremely low redshifts (e.g.\ within the ``hockey-stick'' DE model~\cite{Camarena:2021jlr}). Therefore, we do not expect substantial differences in our work, since the DE models considered are smooth, and the deviations from $\Lambda$CDM in the parameter constraints are small.

Nevertheless, we test this expectation, focusing for concreteness on a specific dataset combination, i.e.\ baseline+SNeIa+SH0ES. In the first approach, which mimics what has been done in our work (despite this specific dataset combination has never been explicitly considered), we treat the SH0ES prior as a prior on $H_0$. In the second approach, we self-consistently include the SH0ES information by using the joint \textit{PantheonPlus}+SH0ES likelihood. The results are reported in Tab.~\ref{tab:H0_prior}: we observe no significant shift in our upper limits on $\sum m_{\nu}$, which tighten by approximately $0.005\,{\text{eV}}$ when using the full likelihood instead of a prior on $H_0$. The constraints on all other cosmological parameters are extremely stable. We conclude that our main results and the overall message of our paper are not significantly affected by the choice of treating the SH0ES information as a prior on $H_0$, strategy which in any case only would only affect those constraints which are artificially driven by the tension between cosmological datasets.

\begin{table*}[!ht]
\centering
\begin{tabular}{|l?cc|cc|}
\hline \hline
& \multicolumn{2}{c|}{$\Lambda$CDM+$\sum m_{\nu}$}
& \multicolumn{2}{c|}{NPDDE+$\sum m_{\nu}$}\\
\cline{2-5}
\textbf{Dataset combination} & $\sum m_{\nu}\,[{\text{eV}}]$ & $B_{\text{NO,IO}}$ & $\sum m_{\nu}\,[{\text{eV}}]$ & $B_{\text{NO,IO}}$\\
\hline
baseline+SNeIa+SH0ES ($H_0$) & $<0.053$ & $28.4$ & $<0.049$ & $32.9$\\
baseline+SNeIa+SH0ES (full likelihood) & $<0.048$ & $56.9$ & $<0.044$ & $64.4$\\
\hline \hline
\end{tabular}
\caption{Impact on the resulting $\sum m_{\nu}$ constraints of different treatments of the SH0ES information, tested on two dataset combinations.}
\label{tab:H0_prior}
\end{table*}

\newpage

\bibliography{Mnu}

\end{document}